\documentclass[%
reprint,
superscriptaddress,
jor,
 amsmath,amssymb,
floatfix,
]{revtex4-1}
\usepackage{amsmath}
\usepackage{marvosym}
\usepackage{graphicx,subfigure}         
\usepackage{amsmath} 
\usepackage{amssymb,stmaryrd}   
\usepackage{setspace}
\usepackage{color}
\usepackage{gensymb}
\usepackage{wrapfig}
\usepackage{mathtools}

\usepackage{flushend}
\usepackage{tikz}

\usepackage{natbib}

\usepackage{comment} 
\usepackage{siunitx}

\usepackage{appendix}
\usepackage{hyperref}
\usepackage{textcomp}

\DeclareUnicodeCharacter{2061}{}

\usepackage{gensymb}

\usepackage{graphicx} 
\usepackage{marginnote}

\hypersetup{colorlinks=true, linkcolor=blue, citecolor=blue, urlcolor=blue,}

\usepackage{ragged2e}
\usepackage{setspace}
\begin{document}

\makeatletter
\renewcommand{\maketitle}{\bgroup\setlength{\parindent}{10pt}
\begin{flushleft}
  \textbf{\@title}

  \@author
\end{flushleft}\egroup
}
\makeatother

\onecolumngrid

\title{\textsf{\huge {Morphogenesis of Spin Cycloids in a Non-collinear Antiferromagnet }}}

\vspace{20cm}
\date{}
\author{%
\textsf{\textbf{Shashank Kumar Ojha$^{1,*}$, Pratap Pal$^{2}$, Sergei Prokhorenko$^{3}$, Sajid Husain$^{4,*}$, Maya Ramesh$^{5}$, Peter Meisenheimer$^{4}$, Darrell G. Schlom$^{5,6,7}$, Paul Stevenson$^{8}$, Lucas Caretta$^{9}$, Yousra Nahas$^{3}$, Lane W. Martin$^{1,10,11}$, Laurent Bellaiche$^{3,12}$, Chang-Beom Eom$^{2}$, Ramamoorthy Ramesh$^{1,4,10,11,13,14*}$\\}
$^{1}$Rice Advanced Materials Institute, Rice University, Houston, TX, 77005, USA\\
$^{2}$Department of Materials Science and Engineering, University of Wisconsin-Madison, Madison, WI 53706, USA\\
$^{3}$Smart Ferroic Materials Center, Physics Department and Institute for Nanoscience and Engineering, University of Arkansas, Fayetteville, Arkansas, USA\\
$^{4}$Department of Materials Science and Engineering, University of California, Berkeley, CA, 94720, USA\\
$^{5}$Department of Materials Science and Engineering, Cornell University, Ithaca, NY, 14853, USA\\
$^{6}$Kavli Institute at Cornell for Nanoscale Science, Cornell University, Ithaca, NY, 14853, USA\\
$^{7}$Leibniz-nstitut fur Kristallzuchtung, Max-Born-Str. 2, 12489, Berlin, Germany\\
$^{8}$Department of Physics, Northeastern University, Boston, MA 02115, USA\\
$^{9}$School of Engineering, Brown University, Providence, RI, USA\\
$^{10}$Department of Materials Science and NanoEngineering, Rice University, Houston, TX, 77005, USA\\
$^{11}$Departments of Chemistry and Physics and Astronomy, Rice University, Houston, TX, 77005, USA\\
$^{12}$Department of Materials Science and Engineering, Tel Aviv University, Ramat Aviv, Tel Aviv 6997801, Israel\\
$^{13}$Materials Sciences Division, Lawrence Berkeley National Laboratory, Berkeley, CA, 94720, USA\\
$^{14}$Department of Physics, University of California, Berkeley, CA, 94720, USA\\
{$^{*}$so37@rice.edu}\\
{$^{*}$rramesh@berkeley.edu}\\
{$^{*}$shusain@berkeley.gov}\\
}}

\date{\today}
\maketitle

\textbf{\large{Pattern formation in spin systems with continuous-rotational symmetry (CRS) provides a powerful platform to study emergent complex magnetic phases and topological defects in condensed-matter physics. However, its understanding and correlation with unconventional magnetic order along with high-resolution nanoscale imaging is challenging. Here, we employ scanning NV magnetometry to unveil the morphogenesis of spin cycloids at both the local and global scales within a single ferroelectric domain of (111)-oriented BiFeO$_3$, which is a non-collinear antiferromagnet, resulting in formation of a glassy labyrinthine pattern. We find that the domains of locally oriented cycloids are interconnected by an array of topological defects and exhibit isotropic energy landscape predicted by first-principles calculations. We propose that the CRS of spin-cycloid propagation directions within the (111) drives the formation of the labyrinthine pattern and the associated topological defects such as antiferromagnetic skyrmions. Unexpectedly, reversing the as-grown ferroelectric polarization from [$\bar{1}$$\bar{1}$$\bar{1}$] to [111] induces a magnetic phase transition, destroying the labyrinthine pattern and producing a deterministic non-volatile non cycloidal, uniformly magnetized state. These findings highlight that (111)-oriented BiFeO$_3$ is not only important for studying the fascinating subject of pattern formation but could also be utilized as an ideal platform for integrating novel topological defects in the field of antiferromagnetic spintronics.\\}}

\setcounter{figure}{0}
\renewcommand{\figurename}{\textbf{Figure}}
 
\twocolumngrid
\section*{{\bf{Introduction}}} 
The emergence of self-organized spatio-temporal patterns in non-equilibrium and homogeneous open systems has long captivated scientists~\cite{Turing1990,Cross:1996p851,walgraef:2012p,cross:2009p,RevModPhys.66.1481}. These patterns, varying in complexity, manifest across a wide range of length scales—from spiraling galaxies to nanoscale Turing patterns in single layers of bismuth~\cite{Fuseya:2021p1031}. In a quest to seek universality, extensive studies on a multitude of physical systems and mathematical models have led to the understanding that symmetry breaking and finite wavelength instabilities are key drivers of such pattern formation~\cite{Cross:1996p851,walgraef:2012p,cross:2009p}. In particular, since instabilities naturally arise near phase transitions, the study of pattern-forming systems has significantly advanced our understanding of phase transitions across various dimensions~\cite{walgraef:2012p,cross:2009p,Zurek1996,Kibble2007}.

Systems with continuous-rotational symmetry (CRS) in two dimensions (2D) are particularly intriguing. In such systems, according to the topological connectivity of the motif, point and line topological defects (regions of locally vanishing order parameters) often emerge as critical elements~\cite{Mermin1979:p591,Lyons1992}. These defects play a crucial role in understanding the dynamics and evolution of patterns and lie at the core of understanding the disorganization of the ordered phase~\cite{Zurek1996,Kibble2007}. For instance, the melting of 2D crystals, as described by the Kosterlitz-Thouless-Halperin-Nelson-Young (KTHNY) theory~\cite{Kosterlitz:1972pL124,Kosterlitz:1973p1181,Halperin:1978p121,Young:1979p1855}, illustrates how various defects such as dislocations and disclinations play pivotal roles in phase transitions.

In the context of magnetism~\cite{PhysRevA.46.7519}, these topological defects assume prime importance, as they give rise to non-trivial swirling spin textures~\cite{Schoenherr:2018p465} such as merons (half-skyrmions) and skyrmions that have shown promise for the realization of ultra-low power electronics~\cite{Fert:2017p17031,amin2023}. Such topological objects have been explored in various condensed-matter systems with broken space inversion symmetry, such as magnets ~\cite{Fert:2017p17031,nagaosa2013topological} ferroelectrics/multiferroics ~\cite{das2019observation,govinden2023ferroelectric} and chiral liquid crystals~\cite{nych2017spontaneous,ackeerman2017}. Although they have been stabilized in magnetic systems with CRS, their integration into magnetoelectric (ME) multiferroic systems remains relatively unexplored~\cite{PhysRevLett.128.187201,Chaudron2024}.
Owing to the inherent ME coupling in such systems, local topological-spin textures could potentially be toggled with an external electric field, opening new possibilities for advanced applications in an electric field controlled antiferromagnetism \cite{Fert:2024p015005}. In this work, we demonstrate how CRS along with suitable electrostatic/geometrical boundary conditions could be used to tailor novel isolated and hybrid topological-spin textures in the multiferroic BiFeO$_3$.

During the last two decades, BiFeO$_3$ has emerged as an exciting platform for the development of energy efficient, spin-based electronics~\cite{Bibes:2008p425,Heron:2014p370,Manipatruni:2019p35,Parsonnet:2022p087601,Fert:2024p015005,Huang2024,husain2024non}. At room temperature, BiFeO$_3$ has a noncentrosymmetric distorted rhombohedral structure~\cite{Moreau1971} (consisting of two corner shared pseudocubic unit cells) and exhibits a large spontaneous ferroelectric polarization {\bf{\textit{P}}} $\simeq$100 $\mu$C/cm$^2$ which can point along any of  the $<$111$>$ (hereafter, all crystallographic directions/planes are described in the pseudocubic notation) (Supplementary Fig. 1). The magnetic structure can be approximated to be a G-type antiferromagnet. The presence of the antisymmetric Dzyalozhinski-Moriya interaction (DMI), however, leads to a small canting that further couples to the polarization to yield a long period ($\sim$ 62 nm) cycloid~\cite{Sosnowska1982,Gross2017,Meisenheimer2024} which rotates in a plane containing {\bf{\textit{P}}} and  {\bf{\textit{k}}}, where {\bf{\textit{k}}} points along a $<$110$>$ and is perpendicular to {\bf{\textit{P}}}~\cite{Burns2020}. The cycloid is accompanied by a spin-density wave perpendicular to the cycloidal plane (Supplementary Fig. 1).

	\begin{figure*}
		\centering{
			{~}\hspace*{-0.2cm}
			\includegraphics[scale=.45]{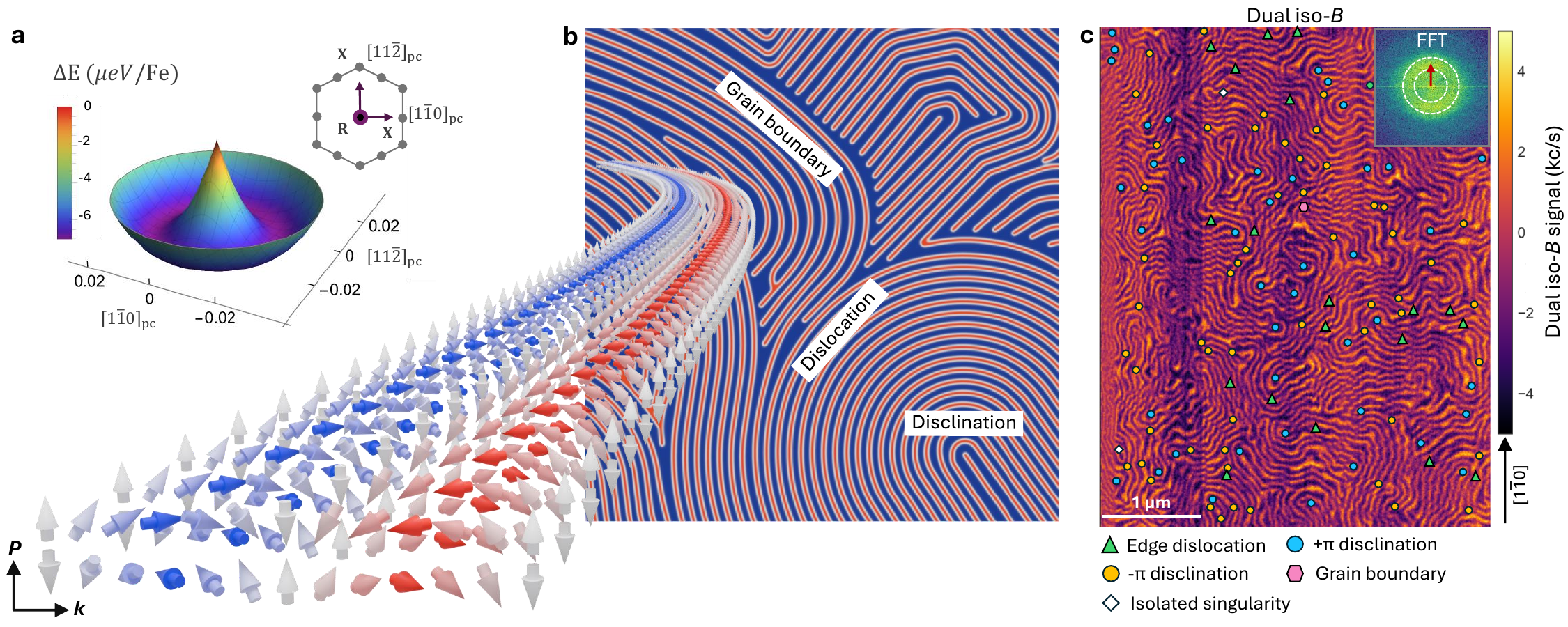}
			\caption{\textbf{Continuous-rotational symmetry and morphogenesis of spin cycloids.} (a) Dependence of the spin cycloid energy on the propagation vector within the (111) calculated using the \emph{ab initio} based Hamiltonian~\cite{Meyer:2023p024403}. The horizontal axes represent the [1-10] and [11-2] orthogonal projections of the cyloid propagation vector centered on the $R$ point of the pseudocubic Brillouin zone (BZ), as shown in the inset. The inset shows the corresponding (111) BZ cross-section with high symmetry $X$ and $R$ points indicated by gray and purple circles, respectively. (b) A schematic to show the continuous rotation of spin cycloids (red and blue stripes) in (111) of BiFeO$_3$. When regions with different local orientational order meet, defects such as disclinations, edge-dislocations and grain boundaries are formed. The actual spatial variation of spins for a curvy stripe is shown projecting outward from the pnael (b). As discussed, spins form a cycloid in the {\bf{\textit{P}}}-{\bf{\textit{k}}} plane which is further modulated by an out-of-plane (perpendicular to {\bf{\textit{P}}}-{\bf{\textit{k}}} plane) tilting due to DM interaction (for more details see Supplementary Fig. 1). Here, two cycloids correspond to two anti-parallel neighbors of Fe sub-lattice.  (c) Magnetic contrast obtained in dual iso-$B$ mode on a 600 nm [111] oriented BiFeO$_3$. Various types of topological defects have been marked with different symbols listed at the bottom. Fast Fourier transform of a 4$\mu$m$\times$4$\mu$m section (from bottom) of the image is shown in the inset. Spread in the $\bm{k}$ vectors is shown with two concentric dotted white circles. A vertical red arrow shows the average $\bm{k}$. This also represents the equiax plane for {\bf{\textit{k}}}-vector in the (111) plane.} \label{fig:1}}
	\end{figure*}

 	\begin{figure*}
		\centering{
			{~}\hspace*{-0.2cm}
			\includegraphics[scale=.5]{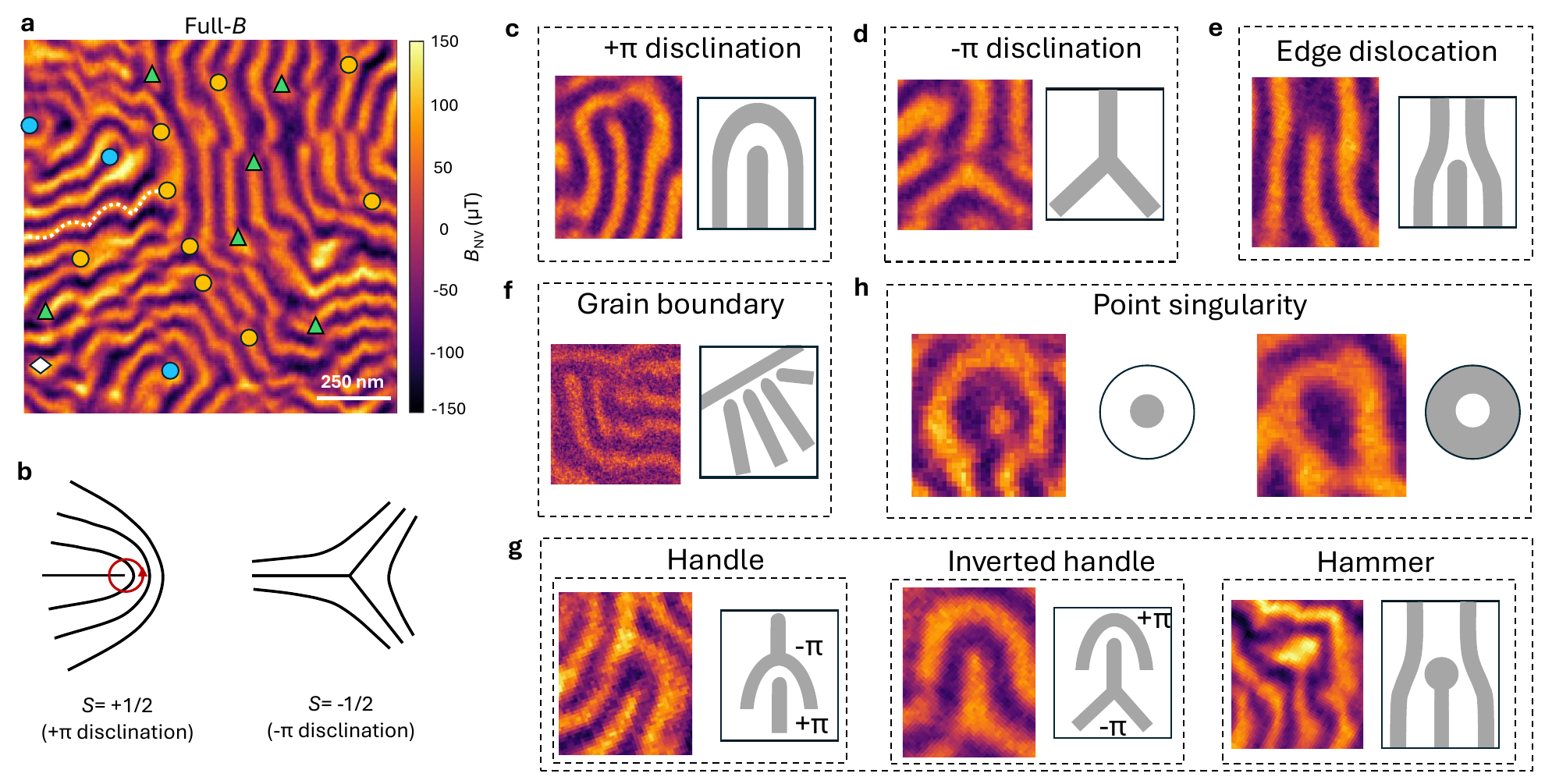}
			\caption{\textbf{Taxonomy of topological defects and interactions among them.} (a) Magnetic contrast obtained in full-$B$ mode on a 300 nm thick (111) BiFeO$_3$ sample. A white dashed wavy line is used to highlight the oscillatory pattern in cycloids. Symbols are used to mark different topological defects. (b) Schematics to show the formation of two types of elementary line defects in planar systems with CRS. Left schematic shows the formation of +$\pi$ disclination. Here the angle $\pi$ comes from the total angle covered by the director field when one encircles around the line defect in anticlockwise direction (shown with red circle). This leads to a winding number $S$=+1/2 and -1/2 (right schematic) for +$\pi$ and -$\pi$ disclinations respectively. (c-f) General class of topological defects observed in homogeneous 2D systems with CRS. (g) Handle, inverted handle and hammer like structure. (h) Isolated point singularity for both spin up (left panel) and down (right panel). Here, all the images have been taken from dual iso-$B$ scans on samples with thickness 300 nm, 600 nm and 1000 nm.} \label{fig:2}}
	\end{figure*}

In the bulk, for each direction of {\bf{\textit{P}}}, symmetry conditions allow for only six propagation vectors '{\bf{\textit{k}}}' related through $\pi/3$ rotations and corresponding to three antiferromagnetic domains~\cite{PhysRevLett.110.217206}. At the same time, recent first principles calculations~\cite{Meyer:2023p024403} indicate that such six-fold anisotropy could be negligible at the length scale of the cycloidal modulation. Particularly, reconstructing the \textit{ab initio} calculated energy surface of antiferromagnetic spin cycloids in monodomain BiFeO$_3$ (Figure \ref{fig:1}a) reveals that the cycloid propagation directions contained in the (111) are energetically degenerate (Methods).

In this context, a critical question emerges: What is the expected spatial pattern of the cycloids on the (111) surface of BiFeO$_3$? A common intuition suggests that the degeneracy in orientational order, imposed by the CRS of spin cycloids, would lead to the continuous rotation of cycloids within the (111) plane (Fig. \ref{fig:1}b). Additionally, regions with different local orientational order could interact and compete, potentially giving rise to topological defects such as disclinations and dislocations (discussed later). This makes (111)-oriented BiFeO$_3$ a rich platform for exploring complex and non-trivial spin textures. In this article, by using high-resolution nitrogen-vacancy diamond-based scanning probe microscopy, we demonstrate how the CRS imposed on the pseudocubic (111) surface of BiFeO$_3$ leads to the complete morphogenesis of the spin structure, resulting in a glassy labyrinthine pattern.~\cite{PhysRevResearch.2.042036}. This pattern features a diverse range of isolated and hybrid topological defects, emphasizing the intricate interplay between symmetry and topological defect formation. Interestingly, electrical switching of the as-grown {\bf{\textit{P}}} direction from [$\bar{1}$$\bar{1}$$\bar{1}$] to [111] eliminates the labyrinthine patterns, resulting in a non cycloidal, uniformly magnetized state that may open new avenues for exploring emergent altermagnetism in BiFeO$_3$ \cite{bernardini2024ruddlesden}.

It is emphasized that having a single ferroelectric domain is crucial for this study as the presence of multiple ferroelectric domains and their associated boundaries would introduce additional boundary conditions, thereby violating the requirement of CRS. In this spirit, single domain BiFeO$_3$ films were grown on SrRuO$_3$-buffered SrTiO$_3$ (111) substrates as a model system (Methods and Supplementary Fig. 2). In this configuration, the rhombohedral distortion axis of BiFeO$_3$ aligns with the normal direction [111] of the SrTiO$_3$ substrate, leading to the stabilization of a single ferroelectric/ferroelastic domain~\cite{Waterfield:2016p177601} (Supplementary Fig. 2). The metallic SrRuO$_3$ bottom electrode effectively screens the depolarizing field in the BiFeO$_3$ film, thereby enforcing the \textbf{\textit{P}} vector to point completely into the film plane (specifically along the [$\bar{1}$$\bar{1}$$\bar{1}$]  in as grown state) resulting in the formation of a single ferroelectric domain~\cite{Waterfield:2016p177601} (Supplementary Fig. 3).


\section*{Results}

{\bf{Real-space imaging of spin cycloids:}} In order to study the nature of the spin cycloids on the (111) surface of BiFeO$_3$, we employed high-resolution nitrogen-vacancy diamond-based scanning probe microscopy (Extended Fig.1 and Methods)~\cite{Maze2008,rondin2014magnetometry}. Figure \ref{fig:1}c shows one representative dual iso-$B$ plot for a 600-nm-thick BiFeO$_3$ film. As is evident, both the spin-up and spin-down stripes that meander within the (111) of BiFeO$_3$ form percolating networks and give rise to a glassy labyrinthine pattern~\cite{PhysRevResearch.2.042036} (additional data, Extended data Fig. 1). 
This glassy behavior is qualitatively consistent with our theoretical calculations (Fig. \ref{fig:1}a) demonstrating that the long-wavelength cycloidal modulations of the G-AFM BiFeO$_3$ state are energetically degenerate for propagation vectors confined in the (111). Such a pattern has previously been observed in ferroelectrics and is generally regarded as an intermediate stage between the ordered and disordered phases, resulting from competing interactions ~\cite{Nahas2020,RevModPhys.95.025001,PhysRevE.66.026203}. As discussed earlier, the formation of any pattern has its roots in a finite wavelength instability of the homogeneous state. In multiferroic systems, this instability is often driven by long-range dipole-dipole interactions or the anti-symmetric DMI. \cite{Li:2023p043109}. The fast Fourier transform (FFT) (inset, Fig. \ref{fig:1}c) of the dual iso-$B$ image appears isotropic and reveals an average wavelength of $\sim$68 nm, close to the cycloid periodicity in the bulk of BiFeO$_3$~\cite{Burns2020}. This observation implies that the formation of such a labyrinthine is connected to the well-known spin-cycloid instability in BiFeO$_3$~\cite{Burns2020}. Since the dual iso-$B$ data is semi-quantitative, to validate our observation, we have also recorded measurements in full-$B$ mode (Fig. \ref{fig:2}a). As seen, cycloid-propagation directions indeed continuously rotate within the  (111), thereby validating our observation of a labyrinthine pattern (Extended data Fig. 2 for additional full-$B$ data). Interestingly, we also found that, unlike the case of (001), the stray field in (111) oriented films can not be solely attributed  to the spin density wave (see Supplementary Note 2 for more details).

Further, apart from the isotropic nature of the FFT (as expected for a labyrinthine pattern), there is also a sparsing nature of the cycloid width (inset, Fig. \ref{fig:1}c, dotted white circles). This would mean that the periodicity of the cycloids has a distribution spanning across all directions within the (111)  (Extended data Fig. 3, 4, 5). We also found that the modulus of the $B$-field also varies randomly across cycloids (Extended data Fig. 4). Since the magnitude of the canted moment is directly linked to the cycloid periodicity and the strength of the local $B$-field~\cite{PhysRevApplied.17.044051}, these observations strongly indicate that spin cycloids have undergone subtle changes even at the unit-cell level.  This observation may be attributed to the coexistence of instabilities in both the spatial and temporal domains~\cite{walgraef:2012p}. In the context of reaction-diffusion systems, it has previously been noted that instabilities in the temporal domain, such as a Hopf instability, could lead to oscillating patterns~\cite{liu2007oscillatory,Mullin1989,Agladze1984}. When these temporal instabilities interact with spatial instabilities, such as a Turing  or cycloidal instability~\cite{Turing1990}, complex patterns could emerge with variations even at local scales. In particular, the presence of aperiodic modulations perpendicular to the cycloid propagation within the samples (dotted white line. Fig. \ref{fig:2}a) suggest that a similar interplay of instabilities is relevant to this system.

{\bf{Topological defects:}} In systems with CRS, the local director field, which describes the direction and degree of orientational order, exhibits infinite degeneracy and therefore can align in any direction within a planar geometry. This leads to the formation of topological-line defects due to competing nature of local orientational order. The two most elementary topological defects, ±$\pi$ disclinations, which are characterized by winding numbers  $S$=±1/2, are illustrated (Fig. \ref{fig:2}b). To explore these in the present case, we have mapped out the discontinuities in local cycloidal order across several samples (Fig. \ref{fig:1}c, Extended data Fig. 1) and found a dense distribution of such disclinations. We have also discovered a variety of hybrid-topological defects (identified by different symbols, Fig. \ref{fig:1}c, Fig. \ref{fig:2}a). A catalog of the same is presented in panels (c)-(g) in Fig. \ref{fig:2}.

 The observed elementary defects can be readily understood using homotopy-theory arguments~\cite{Mermin1979:p591}. The $\pm \pi$ disclinations (Fig. \ref{fig:2} c-d) as well as edge dislocations (Fig. \ref{fig:2} e) are a consequence of the spin-cycloid order parameter space which encompasses two planar rotational symmetries. The first symmetry is related to the continuous rotation of the cycloid-propagation vector within the (111) (Fig. \ref{fig:1} a) and gives rise to disclinations~\cite{Mermin1979:p591} (Fig. \ref{fig:2}c-d). The second rotational symmetry shifts the phase of harmonic-cycloid modulation along a given propagation vector and is responsible for the appearance of edge dislocations~\cite{Chaikin95}.

One particularly interesting aspect of such elementary-topological defects is that they possess a non-zero topological charge (attached to their topology) and can interact with each other \cite{Nagase2021,Nahas2020,Gim2017,Giomi2017}. We observed several hybrid-topological defects arising from such interactions. For example, arrays of disclinations result in grain boundaries (Fig. \ref{fig:2}f) while the interaction between $\pm\pi$ disclinations leads to the formation of structures such as handle structures (Fig. \ref{fig:2}g)~\cite{Nahas2020}. Additionally, a point defect can interact with an edge dislocation to form a hammer-like structure. 

 	\begin{figure*}
		\centering{
			{~}\hspace*{-0.2cm}
			\includegraphics[scale=.55]{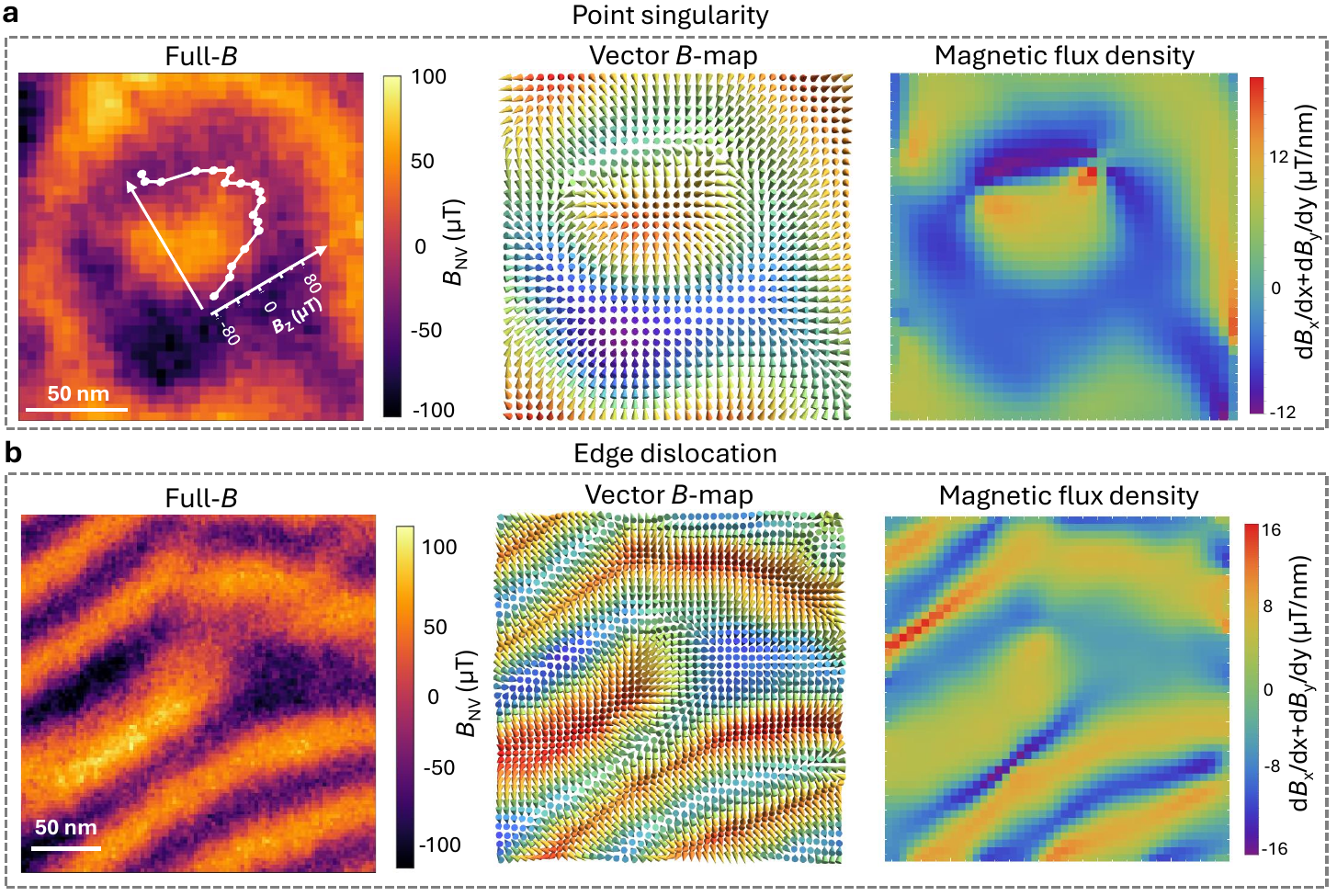}
			\caption{\textbf{Antiferromagnetic skyrmion and nature of spin texture around topological defects.} (a) (Left panel) Full-$B$ image taken around a point singularity. In-set graph shows the line scan of $z$-component of the stray field across the singularity. The corresponding vector map of the normalized $B$-field and magnetic flux density has been shown in middle and right panels respectively.  (b) (Left panel) Full-$B$ image taken around an edge dislocation. The corresponding vector map of the normalized $B$-field and magnetic flux density is shown in middle and right panels respectively. See Extended data Fig. 6 and 7 for additional data.} \label{fig:3}}
	\end{figure*}

Beyond such elementary and hybrid defects, isolated point singularities are also discovered (Fig. \ref{fig:2}h). It is hypothesized that the cycloidal order initially forms locally before expanding across the sample—a phenomenon frequently observed in reaction-diffusion systems~\cite{Turing1990}, which serve as models for pattern formation~\cite{Lyons1992}, and have also been applied to antiferromagnets~\cite{Merle:2019p042111,Zakany:2022p048102,Li:2023p043109}. These singularities are particularly intriguing, as they have the potential to host antiferromagnetic skyrmion-like spin textures. In the present case, they can be visualized by considering the complete winding of spins along the propagation of the underlying spin cycloids, fully modulated by the rotational symmetry in the (111) of BiFeO$_3$. We note that, while ferroelectric solitons (i.e. topologically protected textures) have been crafted in epitaxial superlattices of the form SrTiO$_3/$BiFeO$_3$/SrTiO$_3$ ~\cite{govinden2023ferroelectric}, their observation in magnetic counterparts remains illusive. To investigate this, we constructed a vector map of the local $B$-field around the singularity and found continuous rotation of the $B$ field as expected for a skyrmion (Fig. \ref{fig:3}a). Since there is no direct one-to-one mapping between the $B$-field and the actual spin texture, we focus on only the $z$ component ($B_Z$), which fully preserves the rotational symmetries of the out-of-plane magnetization~\cite{Dovzhenko2018}. We find that its variation across the singularity (inset, left panel of Fig. \ref{fig:3}a) closely resembles the $B_Z$ profile around the skyrmions previously reported in magnetic multilayer systems through NV magnetometry~\cite{Dovzhenko2018}. In a recent theoretical calculation for BiFeO$_3$, it is suggested that spin winding generates an extra polarization in response to the spin-current model: $P_{ME}$$\approx$$e_{ij}$$\times$ ($S_{i}$$\times$ $S_{j}$) parallel to the polarization along the [111], thereby stabilizing the skyrmion in 3D~\cite{Li:2023p043109}. While the evolution of the local $B$-field in the present case strongly attests to the presence of a skyrmion-like spin texture, at least in 2D on the (111) surface, further investigation is needed to fully understand its structure in 3D. Inspired by this, we have also constructed the $B$-field vector map around an edge dislocation (Fig. \ref{fig:3}b) and found that the evolution of the $B$-field is qualitatively consistent with that of a meron (topological spin texture equivalent to half of a skyrmion). These observations emphasize the potential of the (111) surface of BiFeO$_3$ for integrating novel spin textures and their electrical manipulation.
  
	\begin{figure*}
		\centering{
			{~}\hspace*{-0.2cm}
			\includegraphics[scale=.55]{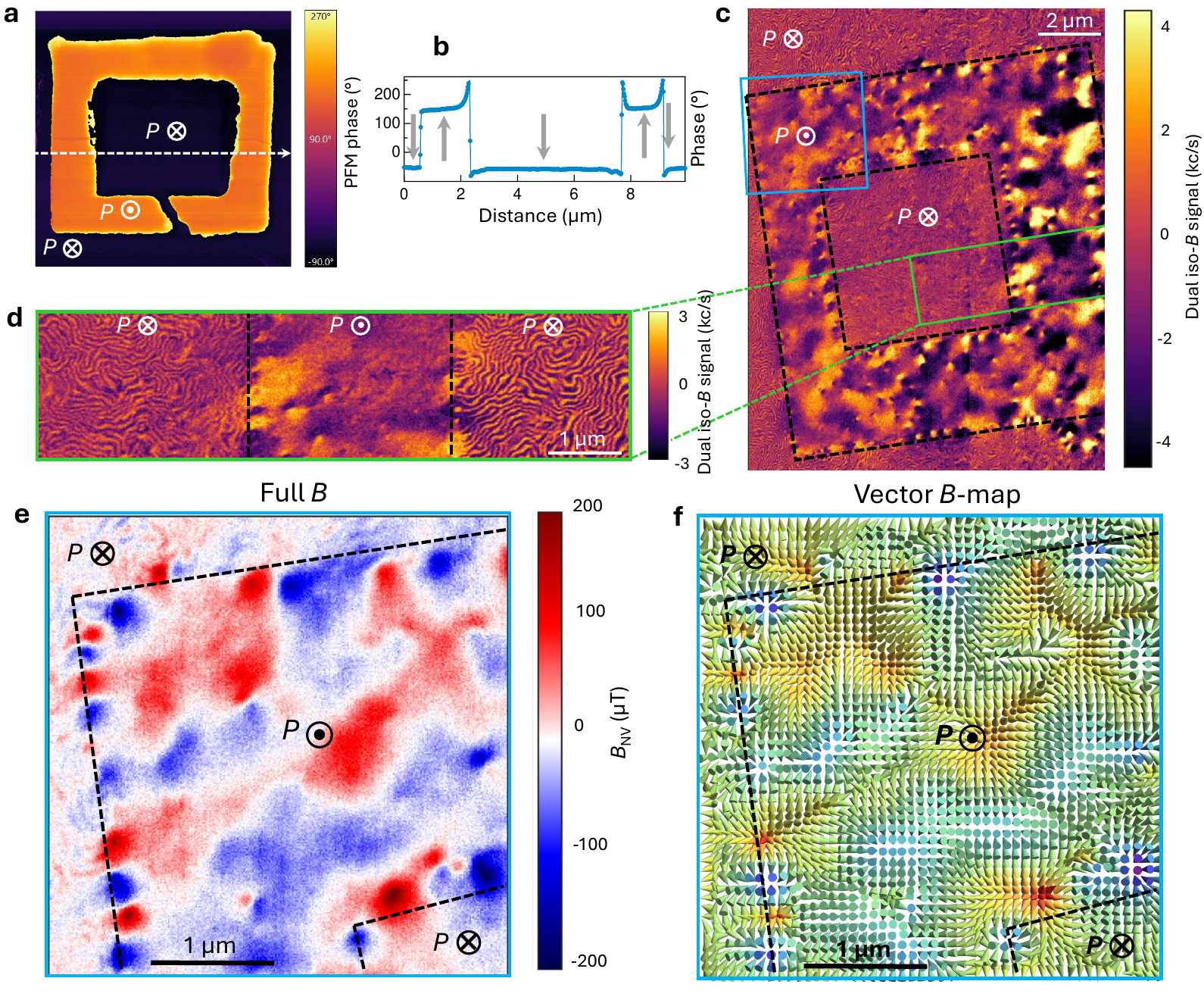}
			\caption{\textbf{Electric field switching.} (a) Piezo-force microscopy (PFM) phase image of a box-in-box pattern switched with a conductive AFM tip on a 1000 nm BiFeO$_3$/SrRuO$_3$/SrTiO$_3$ (111) sample. To create this box-in-box switched structure, first a 8$\mu$m$\times$8$\mu$m region was switched with -40 V and then a 5$\mu$m$\times$5$\mu$m region was switched back with +40 V. For as-grown films, the {\bf{\textit{P}}} vector points into the substrate along [$\bar{1}$$\bar{1}$$\bar{1}$]. Upon application of a -40 V, the polarization switches by 180 degree which can be again switched back to original by reversing the voltage to +40 V. (b) The PFM phase  along a horizontal line through the box-in-box pattern (see Supplementary Fig. 3 for more details). (c) Corresponding dual iso-$B$ image of box-in-box pattern written by PFM. Black dotted lines denote the switched boxes. Blue and green boxes denote two regions where high resolution dual iso-$B$ and full-$B$ images were taken, presented in panels (d) and (e) respectively. Vector map of the normalized $B$-field for panel (e) has been shown in panel (f). Un-normalized vector $B$ map is shown in the Extended data Fig. 9. Also, see Extended data Fig. 8  for $x$, $y$ and $z$ components of $B$-field used to construct vector $B$ map shown in the panel (e).  Additional data set on upward poled region on another sample is shown in Extended data Fig. 10.} \label{fig:4}}
	\end{figure*}


 
{\color{black}{\bf{Electric-field switching:}}} Our initial premise from the data (Figs. \ref{fig:1}-\ref{fig:3}) was that such glassy labyrinthine patterns arise as a consequence of competing interactions during the emergence of the cycloid structure upon cooling from the synthesis temperature. We, therefore, hypothesized that electrical switching of the ferroelectric order in local regions could possibly provide a pathway to increase the degree of order in the cycloid (effectively "electrical annealing") and perhaps could even lead to a perfectly hexagonally ordered cycloid lattice~\cite{PhysRevLett.110.217206}.  In turn, bipolar electric-field switching of the ferroelectric state were carried out using piezoresponse force microscopy (PFM) (Methods and Figure \ref{fig:4}a) to locally switch the BiFeO$_3$ layer (in a box-in-box pattern) as well as switching through a platinum top electrode in a metal-ferroelectric-metal capacitor structure (Extended data Fig. 11). Fig. \ref{fig:4}a illustrates the switching of the polarization (which, in the as-grown state, is along [$\bar{1}$$\bar{1}$$\bar{1}$], i.e., towards the bottom SRO electrode) to point along [111], as evidenced by the change in the phase of the piezoresponse signal by 180$^\circ$ (Fig. \ref{fig:4}b). We were surprised to find that the first switching event (from the down- to up-poled state) led to the destruction of the cycloid state at the surface and the emergence of a non cycloidal, uniformly magnetized state. Switching back (from the up- to down-poled state), led to the re-emergence of the cycloid state. This deterministic switching is captured in the dual iso-$B$ image which is qualitatively related to the $B$-field emerging from the sample (Fig. \ref{fig:4}c), and a higher resolution/magnification image is also provided (Fig. \ref{fig:4}d) along with the full-$B$ map from a portion of this doubly switched region (Fig. \ref{fig:4}e) with the corresponding vector $B$ map (Fig. \ref{fig:4}f). The effective $x$, $y$, and $z$ components of the $B$-field distribution in the up-poled region does not display the cycloid-based labrynthine pattern (Extended data Fig. 8); instead the contrast can be interpreted as non cycloidal, uniformly magnetized state.  Such a uniform, local canted moment structure for BiFeO$_3$ has been theoretically predicted to be an altermagnet phase  \cite{bernardini2024ruddlesden}. Thus, one can visualize such an electric-field-driven change in the cycloid pattern as a magnetic-phase transition on the (111) BiFeO$_3$ surface, conveying a special version of the ME effect. The fundamental origins of such a phase transition are still unclear; however, it is hypothesized that it may be related to flexoelectric/flexomagnetic effects in BiFeO$_3$, as has been illustrated previously \cite{Jeon2013,Yang2017}. 

{\color{black}{\bf{Conclusions and outlook:}}} In summary, we have elucidated how the CRS of cycloid propagation in the (111) drives the comprehensive spatial organization of the spin cycloid, culminating in a diverse spectrum of isolated- and hybrid-topological defects which carry substantial implications for the advancement of antiferromagnetic spintronics. Unexpectedly, we observe that switching of the polarization in such a single domain (111)-oriented BiFeO$_3$ leads to the unwinding of the cycloid into a non cycloidal,uniformly magnetized state over a micron length scale.  Future investigations will be directed towards systematic, temperature-dependent studies to delve deeper into the dynamics of these topological defects~\cite{Dussaux2016} to unravel the underlying universality class of the spin-cycloid phase transition leading to the labyrinthine pattern. Recent breakthroughs in non-local transport on BiFeO$_3$ have highlighted the critical role of the spin cycloid in the conceptualization and realization of magnonic devices~\cite{Meisenheimer:2024p2404639,husain2024non}. We believe that our identification of various spatio-temporal instabilities presents key insights that could inform the development of an exhaustive pattern-forming model for BiFeO$_3$. A detailed understanding of these instabilities will be instrumental in the meticulous tuning of material growth parameters, essential for the creation of a perfect macroscopic 1D spin channel which would open new avenues for electrical controlled antiferromagnetic spintronics.

\bibliographystyle{naturemag}
	\bibliography{111_turing_BFO}

	\noindent\section*{Methods}
	\noindent\textbf{Sample Growth}:  (111) monodomain (single ferroelastic and ferroelectric) BiFeO$_3$ thin films were grown on cubic SrTiO$_3$ (111) substrate, buffered by a thin SrRuO$_3$ ($\sim$ 25 nm) layer, by RF magnetron sputtering. First, a $\sim$ 25 nm thick SrRuO$_3$ bottom electrode layer is deposited by 90$^\circ$-off-axis sputtering at 600$^\circ$C  which is followed by a growth of 300 nm - 1000 nm BiFeO$_3$ films by double-gun off-axis sputtering at 750$^\circ$C with Ar:O$_2$ ratio of 4:1 at a total pressure of 400 mTorr ~\cite{Saenrang2017}. The SrRuO$_3$ layer serves three key functions: first, it acts as a bottom or ground electrode for out-of-plane device measurements; second, it generates a depolarization field that aligns the polarization of the BiFeO$_3$ downward; and third, it serves as a buffer layer to accommodate the large lattice mismatch between the BiFeO$_3$ and the substrate~\cite{Baek2010}. The BiFeO$_3$ target used contains 5 $\%$ excess Bi$_2$O$_3$ to compensate for bismuth volatility during thin film deposition. Before growth, SrTiO$_3$ (111) was first etched for 30 seconds in buffered-hydrofluoric acid and then treated at 900$^\circ$ C for 6 hours under 30 sccm oxygen gas flow. The 100 nm thick BiFeO$_3$ (001)-oriented samples were grown on TbScO$_3$ substrates via reactive molecular-beam epitaxy in a VEECO GEN10 system, using a gas mixture of 80 \% distilled ozone and 20 \% oxygen. The substrate temperature was maintained at 675$^\circ$C during the growth at a chamber pressure of 5 × 10$^{-6}$ Torr.

 \vspace{0.2cm}

	\noindent\textbf{NV magnetometry}:  NV images were obtained using a commercially available Qnami Quantum Microscope-ProteusQ$^{TM}$. Parabolic tapered Quantilever$^{TM}$ MX+ diamond tips were chosen for their superior signal-to-noise ratio and photon collection efficiency, making them well-suited for detecting the very small stray fields present in BiFeO$_3$. Here, a nitrogen defect along with the nearest neighboring vacancy in negative charge state (NV$^-$) is utilized as a single atom quantum sensor by making use of its spin triplet state ($m_s$=0,$\mp$1). In our setup, an external magnet is used to break the degeneracy of the $m_s$=$\mp$1 states and a microwave (MW) source frequency is swept to obtain electron spin resonance (ESR) (between $m_s$=0 to $m_s$=$\mp$1 states) which is then detected optically by measuring photoluminescence (PL) intensity. When the NV diamond tip is scanned through the sample surface, the local stray magnetic field ($B$) projected along the NV$^-$ axis moves the ESR spectra which is used to track the local magnetic contrast. Here we have performed imaging in two modes namely dual iso-$B$ and full-$B$~\cite{Gross2017}. In the former, PL is measured at two MW frequencies near the FWHM of the ESR spectra and its difference (PL($\nu_2$)-PL($\nu_1$)) is used to generate the real space magnetic contrast. In the later, full ESR spectra is fitted at each point and strength of local magnetic field is estimated quantitatively. To calibrate the orientation of NV$^-$ center, a known ferromagnetic sample Ta(2 nm)/MgO/CoFeB(0.9 nm)/Ta(5 nm)  with perpendicular magnetic anisotropy was used. For more details, we refer to the Supplementary note 2.

 \vspace{0.2cm}
	
	\noindent\textbf{PFM measurement}: Piezo force microscopy (PFM) imaging was performed using a Jupiter XR Atomic Force Microscope from Oxford Instruments (Asylum Research). All images were acquired in Dual AC Resonance Tracking (DART) mode, which significantly reduces crosstalk caused by shifts in resonant frequency. This is achieved by tracking the contact resonance frequency and, through a feedback loop, adjusting the drive frequency of the cantilever to maintain resonance. Electric field switching experiments were performed by grounding the bottom SrRuO$_3$ electrode with silver paint (Ted Pella), while a voltage was applied to the top of BiFeO$_3$ film using a Pt/Ir-coated conductive tip from Oxford Instruments.

  \vspace{0.2cm}

  	\noindent\textbf{PE loop measurements on metal-ferroelectric-metal capacitor}: Macroscopic polarization versus electric field measurements were conducted on metal-ferroelectric-metal (M-FE-M) capacitors using a Precision Multiferroic Ferroelectric Tester from Radiant Technologies Inc. The SrRuO$_3$ layer was used as the bottom electrode, while a platinum layer ($\sim$10 nm thick deposited by sputtering at 300K), patterned into a disc shape, served as the top electrode. For NV measurements on region switched under electrode, a gold (Au) top electrode was used as it could be easily dissolved with an appropriate chemical solution. In the current work, a solution prepared by mixing KI and I$_2$  in H$_2$O was used as solvent.

 \vspace{0.2cm}
	{{ 
	\noindent\textbf{Theory}: The dependence of the spin cycloid energy on the propagation vector (Fig. \ref{fig:1}a) was computed using the classical spin Hamiltonian of Ref.~\cite{Meyer:2023p024403}. The Hamiltonian 
 \begin{equation}
    H= -\sum_{i,j} J_{ij} \mathbf{S}_i\cdot \mathbf{S}_j -\sum_{i,j} \mathbf{D}_{ij} \cdot (\mathbf{S}_i\times \mathbf{S}_j)-K\sum_i (S_i^{\mathbf{u}})^2
    \label{eq:Hamiltonian}
 \end{equation}
    includes the symmetric spin exchange interaction up to seventh nearest neighbors ($J_{ij}$), the Dzyaloshinskii-Moriya (DM) interaction up to third nearest neighbors ($\bf{D}_{ij}$) and the easy plane anisotropy ($K$) term. The values of the interaction constants are fitted to density functional theory calculations for the monodomain $R3c$ phase of BiFeO$_3$~\cite{Meyer:2023p024403}. The DM vectors are given by the converse spin-current model $\mathbf{D}_{ij}=(\mathbf{u}\times\mathbf{e}_{ij})$, where $\mathbf{u}$ denotes the unit vector oriented along the pseudocubic $[\bar{1}\bar{1}\bar{1}]$ axis and 
    $\mathbf{e}_{ij}$ are unit vectors connecting spins at sites $i$ and $j$. $S_i^{\mathbf{u}}$ denotes the $[\bar{1}\bar{1}\bar{1}]$ projection of the spin.

    The energy is computed assuming the following spin cycloid ansatz
    \begin{equation}
        \mathbf{S}_i = \cos ((\mathbf{Q}_R+\mathbf{q})\cdot \mathbf{r}_i) \mathbf{u} + \sin ((\mathbf{Q}_R+\mathbf{q})\cdot \mathbf{r}_i) \mathbf{e}_{\mathbf{q}}, 
        \label{eq:ansatz}
    \end{equation}
    where $\mathbf{Q}_R$ denotes the $R$ point coordinates of the pseudocubic BZ while $\mathbf{e}_{\mathbf{q}}$ stands for the unit vector in the direction of the cycloid propagation vector $\mathbf{q}$. 
    An analytical energy expression $E(\mathbf{q})$ is obtained by substituting Eq.(\ref{eq:ansatz}) into Eq.(\ref{eq:Hamiltonian}). Fig. \ref{fig:1}a shows the plot of the energy difference $\Delta E = E(\mathbf{q})-E(\mathbf{0})$ between the energies of the AFM spin cycloid ($\mathbf{q}\neq\mathbf{0}$) and G-AFM states ($\mathbf{q}=\mathbf{0}$) for propagation vector $\mathbf{q}$ lying within the (111) plane.
}}

  \vspace{0.2cm}

\hfill \break
\noindent\textbf{\large{Data availability}}\\	
	The data that support the findings of this work are available from the corresponding authors upon reasonable request.
	
\hfill \break
\noindent\textbf{\large{Acknowledgements}}\\
S.K.O is supported by the NSF-FUSE project. S.H. is supported by the U.S. Department of Energy, Office of Science, Office of Basic Energy Sciences, Materials Sciences and Engineering Division under Contract No. DE-AC02-05-CH11231 (Codesign of Ultra-Low-Voltage Beyond CMOS Microelectronics) for the development of materials for low-power microelectronics. L.W.M. and R.R.  acknowledge that this research was sponsored by the Army Research Laboratory as part of the Collaborative for Hierarchical Agile and Responsive Materials (CHARM)  under Cooperative Agreement Number W911NF-24-2-0100. M.R. and D.S. were funded by ARO-MURI (ETHOS) program (Grant No. W911NF-21-2-0162). S.P., Y.N. and L.B. also thank the Vannevar Bush Faculty Fellowship (VBFF) Grant No. N00014-20-1–2834 from the Department of Defense and the MonArk NSF Quantum Foundry supported by the National Science Foundation Q-AMASE-i Program under NSF Award No. DMR-1906383. The views and conclusions contained in this document are those of the authors and should not be interpreted as representing the official policies, either expressed or implied, of the Army Research Laboratory or the U.S. Government. The U.S. Government is authorized to reproduce and distribute reprints for Government purposes notwithstanding any copyright notation herein. The theoretical calculations were performed at the Arkansas High Performance Computing Center (AHPCC). C.B.E. acknowledges support for this research through a Vannevar Bush Faculty Fellowship (ONR N00014-20-1-2844), the Gordon and Betty Moore Foundation’s EPiQS Initiative, Grant GBMF9065. Ferroelectric and structural measurement at the University of Wisconsin–Madison was supported by the US Department of Energy (DOE), Office of Science, Office of Basic Energy Sciences (BES), under award number DE-FG02-06ER46327.

\hfill \break
\noindent\textbf{\large{Author contribution}}\\	

S.K.O., R.R., P.P. and C.B.E. conceived the idea. P.P. performed thin-film growth, fabricated M-FE-M capacitor structure and performed PE loop measurements under the supervision of C.B.E.. S.K.O. carried out NV magnetometry and PFM measurements under the supervision of R.R. and S.H.. S.P., Y.N. and L.B. performed all the theoretical calculations. P.M. helped with the image processing. M.R. prepared the control sample under the supervision of D.S.. P.S. helped calibrate the orientation of NV$^-$ center. S.K.O. wrote the manuscript with the inputs from R.R., S.H. and S.P.. All authors discussed the results and commented on the manuscript.
	
	\section*{Competing interests}
	The authors declare no competing interests.
	
\clearpage
\newpage

\onecolumngrid

\section*{Extended Data}
\textbf{\large{Extended data for "Morphogenesis of Spin Cycloids in a Non-collinear Antiferromagnet"}}

\setcounter{figure}{0}
\renewcommand{\figurename}{\textbf{Extended Data Figure}}

\vspace{1cm}

 	\begin{figure}[htp]
		\centering{
			{~}\hspace*{-0.2cm}
			\includegraphics[scale=.66]{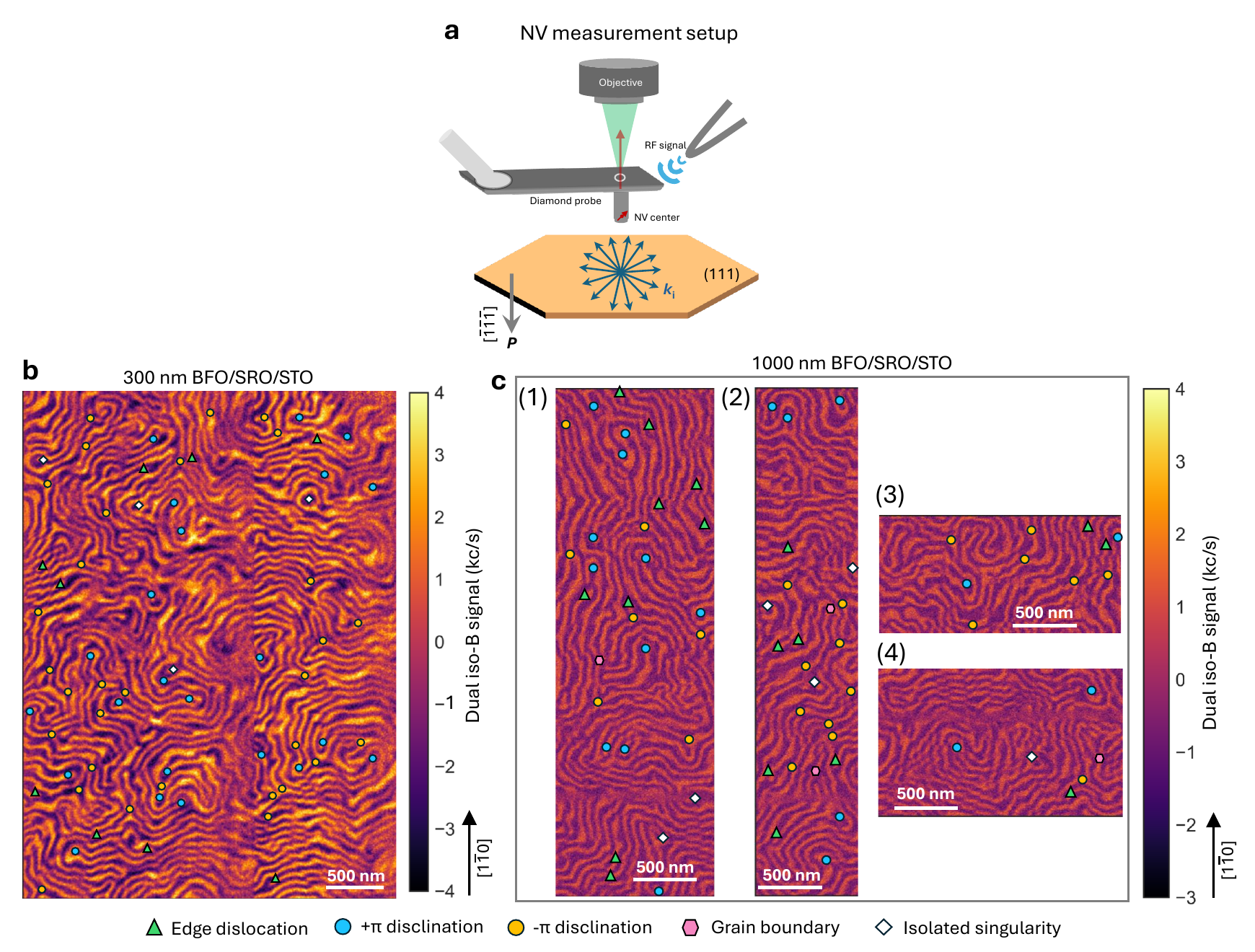}  \caption{(a) Schematic of the nitrogen-vacancy diamond-based scanning probe magnetometry used to image spin cycloids on (111) surface of BiFeO$_3$ films. A representative (111) surface of BiFeO$_3$ being imaged is shown in the bottom panel. [$\bar{1}$$\bar{1}$$\bar{1}$] is the direction of the spontaneous polarization  in the as-grown {\bf{\textit{P}}}-state for our system. Light blue arrows on (111) plane represent the uquiaxial $k$-vector of the spin cycloid within the (111) plane.  Additional dual iso-$B$ images taken on 300 nm (panel (b)) and 1000 nm (panel (c)) thick BiFeO$_3$ samples. Distribution of different types of topological defects have been marked with different symbols as done in the main text.} \label{fig:s4}}
	\end{figure}

\clearpage

 	\begin{figure}[htp]
		\centering{
			{~}\hspace*{-0.2cm}
			\includegraphics[scale=.5]{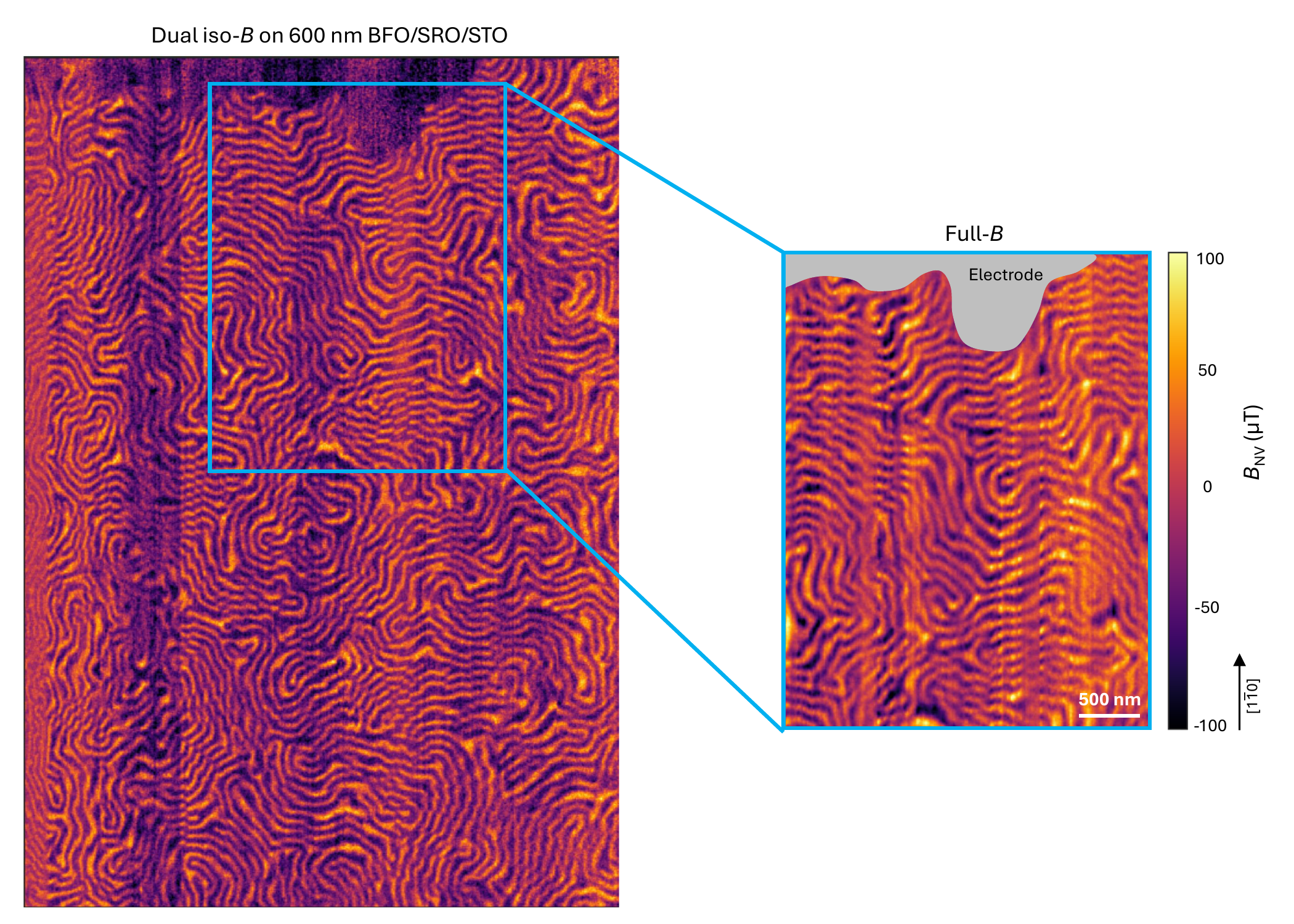}  \caption{ Additional full-$B$ image taken on 600 nm thick BiFeO$_3$ film. Top region covered with slate gray color is the region with platinum electrode.} \label{fig:s5}}
	\end{figure}

\clearpage

  	\begin{figure}[htp]
		\centering{
			{~}\hspace*{-0.2cm}
			\includegraphics[scale=.55]{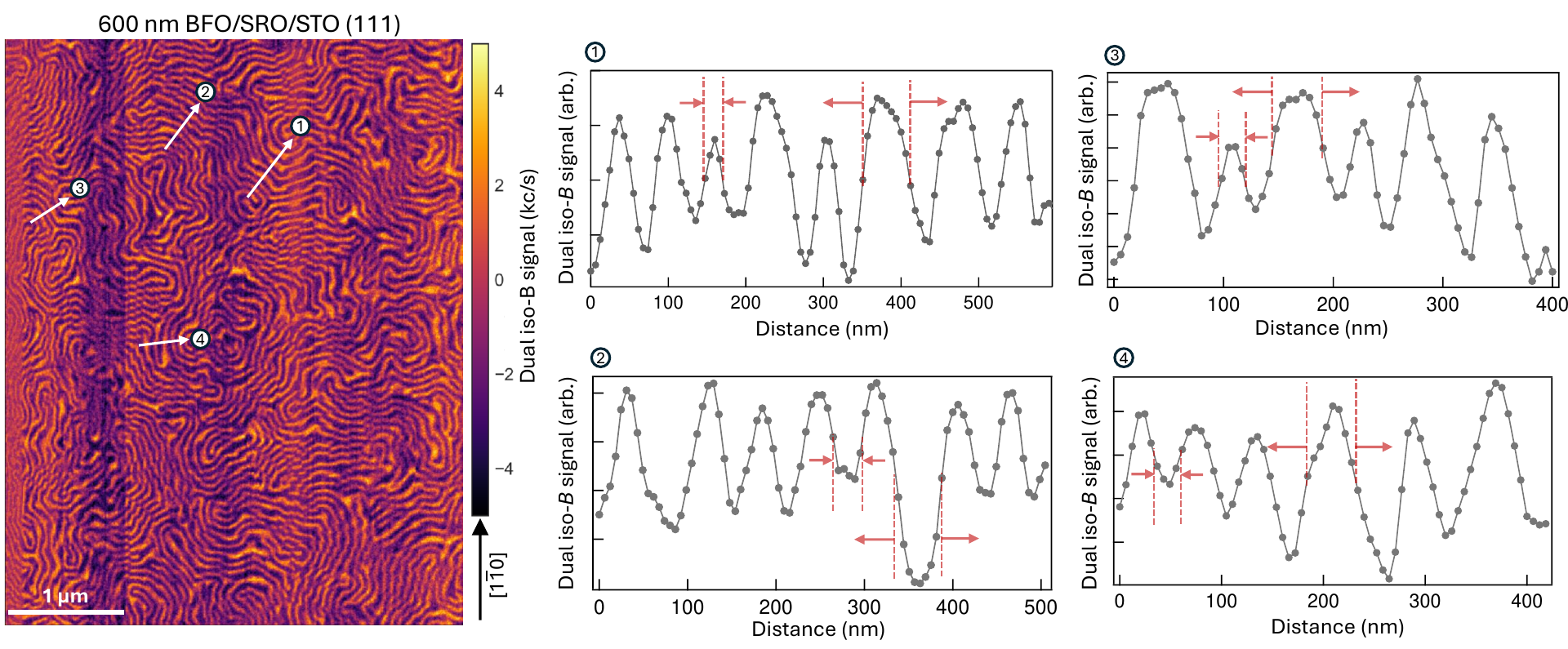}  \caption{Line scans taken along four random directions shown with white arrows. Here, we have deliberately chosen only those regions which have coherent local cycloidal order. Panels numbered 1-4 (in right side) show the corresponding evolution of cycloids along line-scans. As evident, there are variations in the periodicity of cycloids demonstrating morphogenesis even at local scales.} \label{fig:s6}}
	\end{figure}

   	\begin{figure}[htp]
		\centering{
			{~}\hspace*{-0.2cm}
			\includegraphics[scale=.75]{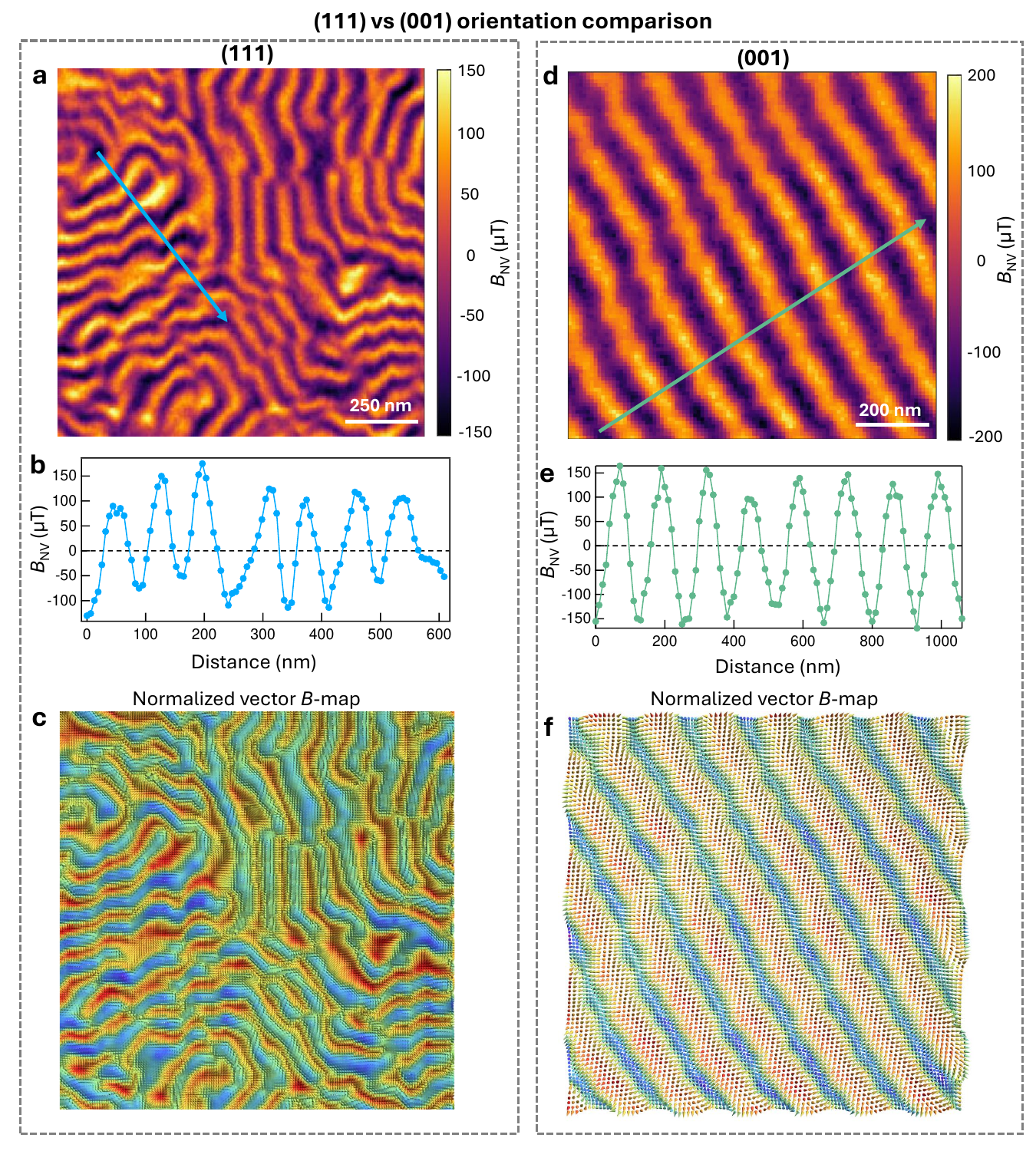}  \caption{ \textbf{Comparison between (111) and (001) oriented BiFeO$_3$ films:} To validate the variation of cycloid periodicity inferred from the analysis of dual iso $B$ signal (shown in the Extended data Fig. 3), we have also analyzed the Full-$B$ data (panel (a)). In panel (b) we present the variation in the magnitude of the stray magnetic field taken along a line scan (shown with the sky blue color arrow in the panel (a)) As is evident, there is drastic variation  in the cycloid periodicity thereby validating our claim. To gain further insight, we constructed stray $B$-field vector maps of spin cycloids, leveraging the linear dependence of $B$-field components in Fourier space~\cite{Dovzhenko:2018p2712,PhysRevApplied.14.024076,PhysRevApplied.17.044051}. This allows reconstruction of the full vector map from a single measurement along the axis parallel to the NV$^-$ center (Supplementary Note 1). In the panel (c), we display the normalized vector map of the same region (panel (a)). As is evident, there is variation not only in the width of the cycloids but also in the magnitude of local stray field, which changes from one cycloid to another. To demonstrate that this is a unique feature of the (111) surface, we have also recorded a full-$B$ measurement on a 100 nm thick (001) oriented BiFeO$_3$ grown on TbScO$_3$ substrate (panel (d)). (e) A line scan taken along a green arrow (shown in the panel (d)) exhibits the homogeneous width of the cycloid. For qualitative comparison, in  panel (f) we also show a similar normalized $B$-field vector map of the same region shown in panel (d). This film exhibits nearly ordered cycloids, further highlighting the critical role of degeneracy in the cycloid propagation vector in driving the spatial organization of cycloids in (111)-oriented BiFeO$_3$. } \label{fig:s6}}
	\end{figure}

 \clearpage

  	\begin{figure}[htp]
		\centering{
			{~}\hspace*{-0.2cm}
			\includegraphics[scale=.55]{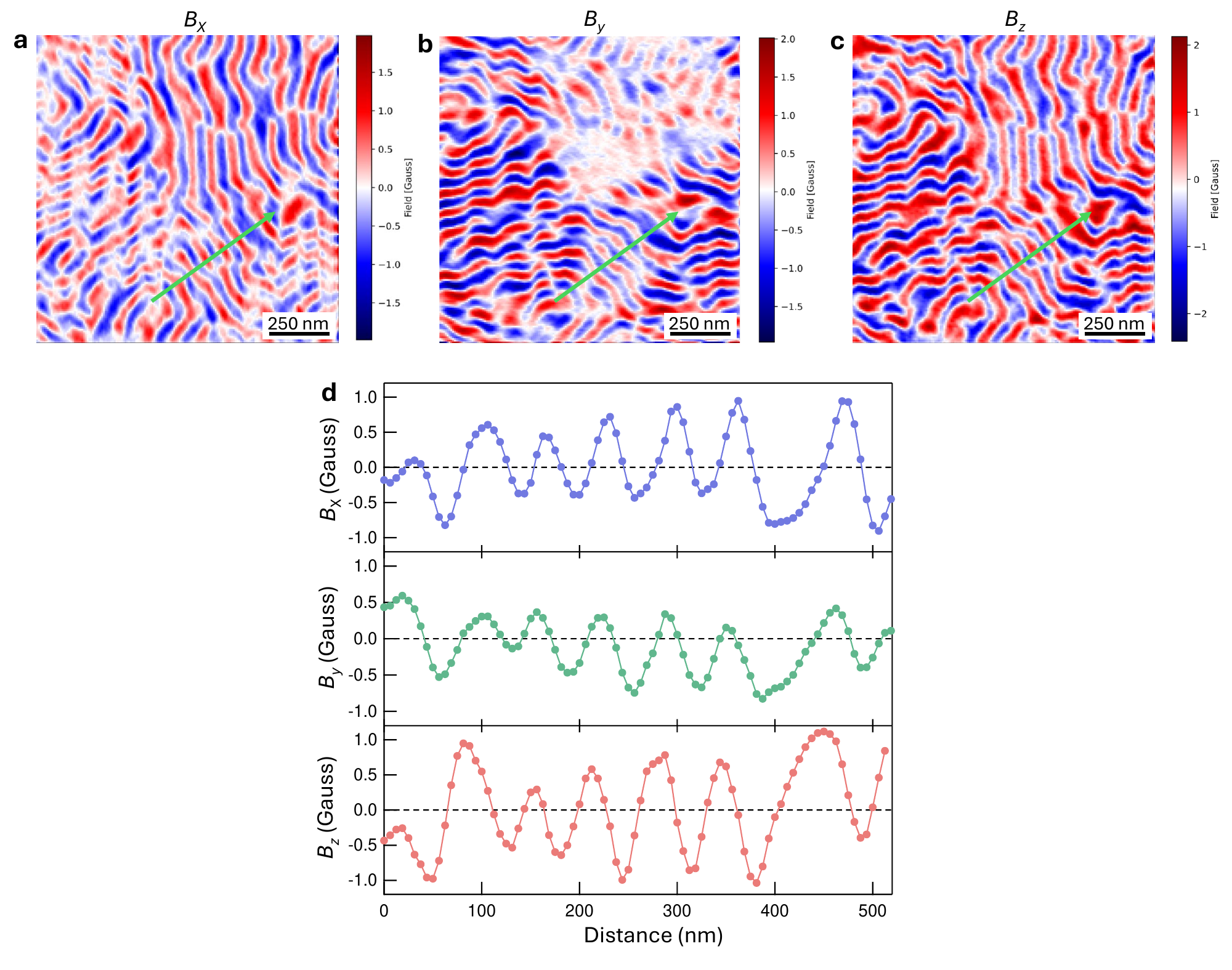}  \caption{\textbf{Full-$B$ reconstruction of the region shown in the panel (a) of Extended data Fig. 4:} (a-c) $x$, $y$ and $z$ components of the stray magnetic field reconstructed in the region shown in the panel (a) of Extended data Fig. 4. Green line shows the direction along which a line scan was taken to show the spatial modulation of the individual $x$, $y$ and $z$ componenets (panel (d)) of the stray magnetic field. } \label{fig:s7}}
	\end{figure}
 
 \clearpage

   	\begin{figure}[htp]
		\centering{
			{~}\hspace*{-0.2cm}
			\includegraphics[scale=.80]{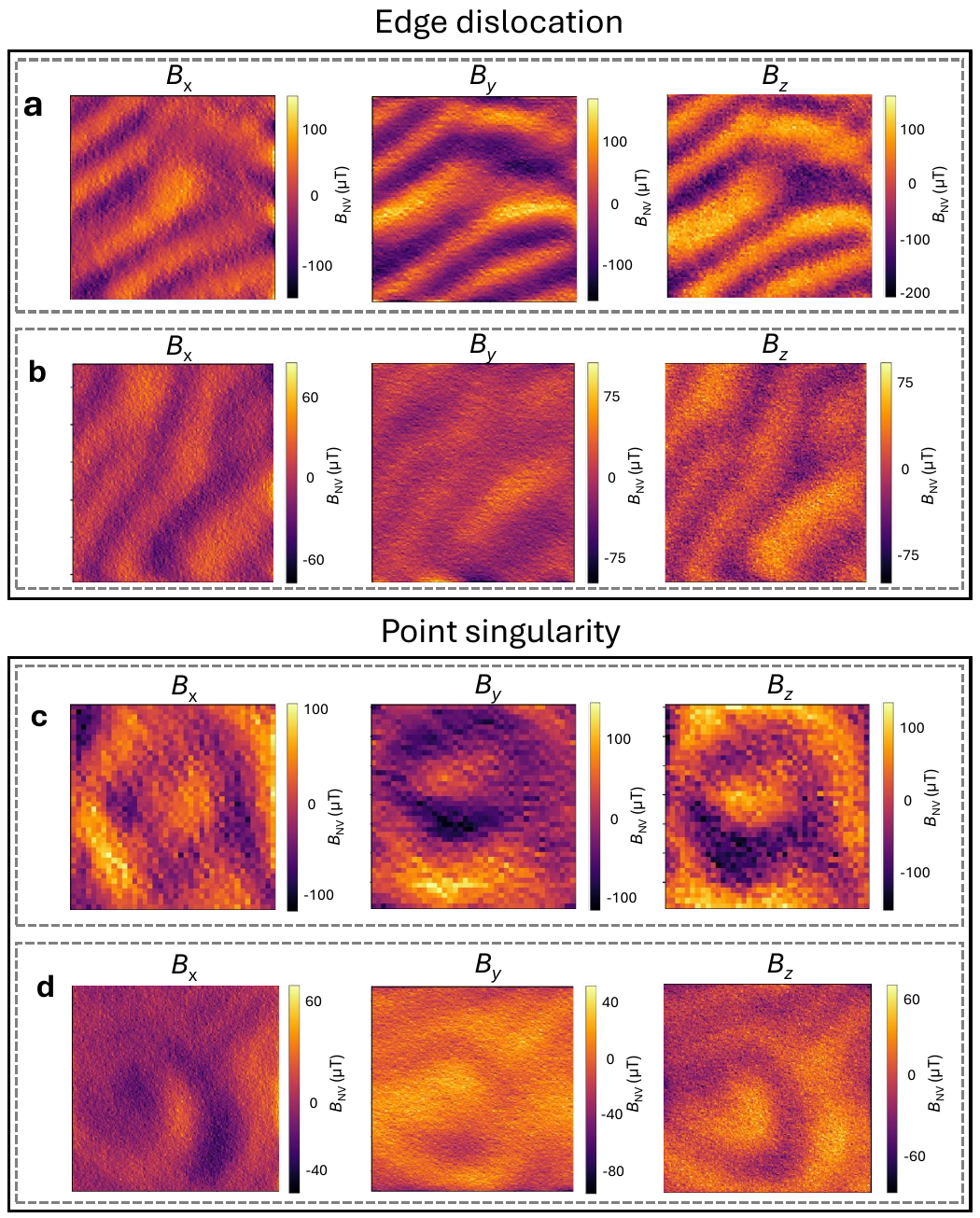}  \caption{ \textbf{Full-$B$ reconstruction around topological defects:} (a-b) $x$, $y$ and $z$ components of the stray magnetic field reconstructed around two edge dislocations. Vector $B$-map for panel (a) has been shown in the Fig. 3a of the main text. The vector map for other one has been shown in the Extended data Fig. 7a. (c-d) $x$, $y$ and $z$ components of the stray magnetic field reconstructed around two isolated spots. Vector $B$-map for panel (c) has been shown in the Fig. 3b of the main text. The vector map for other one has been shown in the Extended Fig. 7b. } \label{fig:s9}}
	\end{figure}
 
\clearpage

   	\begin{figure}[htp]
		\centering{
			{~}\hspace*{-0.2cm}
			\includegraphics[scale=.60]{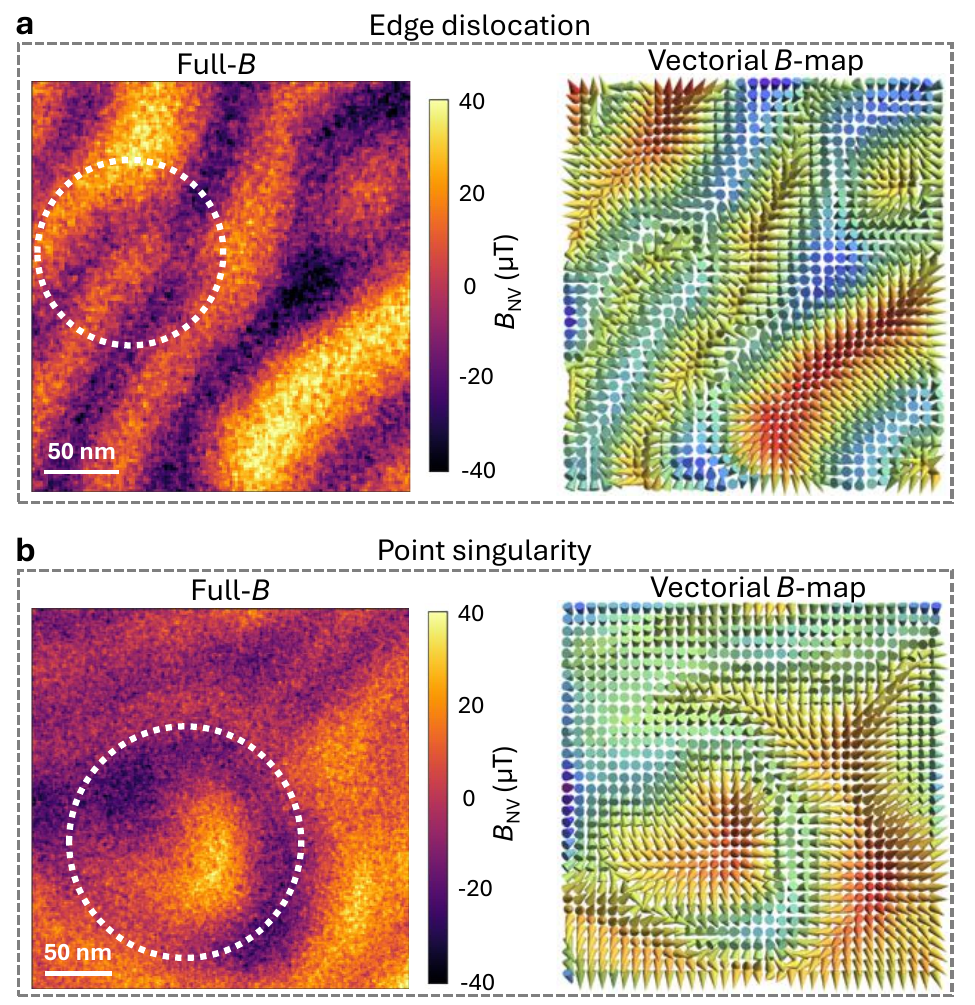}  \caption{ \textbf{Additional $B$ field vector maps near an edge dislocation and point singularity:} For reproducibility, we have mapped vector $B$-maps on another point like singularity and and an edge dislocation. Left panel of (a) shows a Full-$B$ image taken around another edge dislocation indicated with a white dotted circle. The corresponding vector map of the $B$-field has been shown in the right panel. (b) (Left panel) Full-$B$ image taken around a point singularity shown with a white dotted circle. The corresponding vector map of the $B$-field has been shown in the right panel. } \label{fig:s10}}
	\end{figure}

   	\begin{figure}[htp]
		\centering{
			{~}\hspace*{-0.2cm}
			\includegraphics[scale=.65]{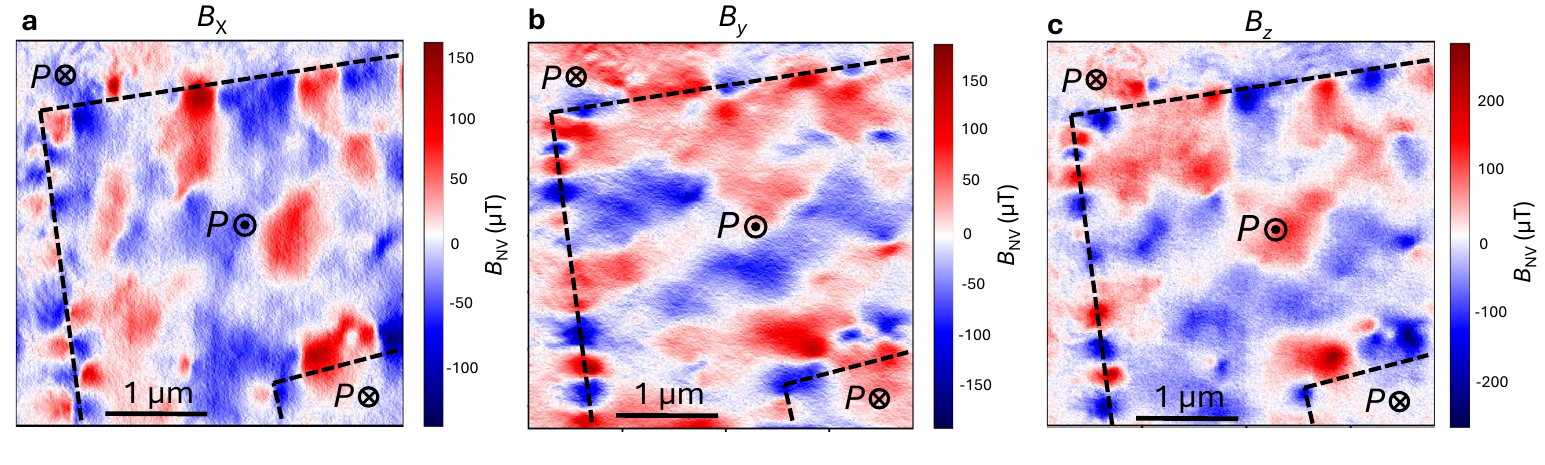}  \caption{ \textbf{Full-$B$ reconstruction in upward poled region shown in the Fig. 4(e) of the main text:} (a-c) $x$, $y$ and $z$ components of the stray magnetic field reconstructed in the upward poled region shown in the Fig. 4e of the main text.}  \label{fig:s11}}
	\end{figure}

\clearpage

   	\begin{figure}[htp]
		\centering{
			{~}\hspace*{-0.2cm}
			\includegraphics[scale=.85]{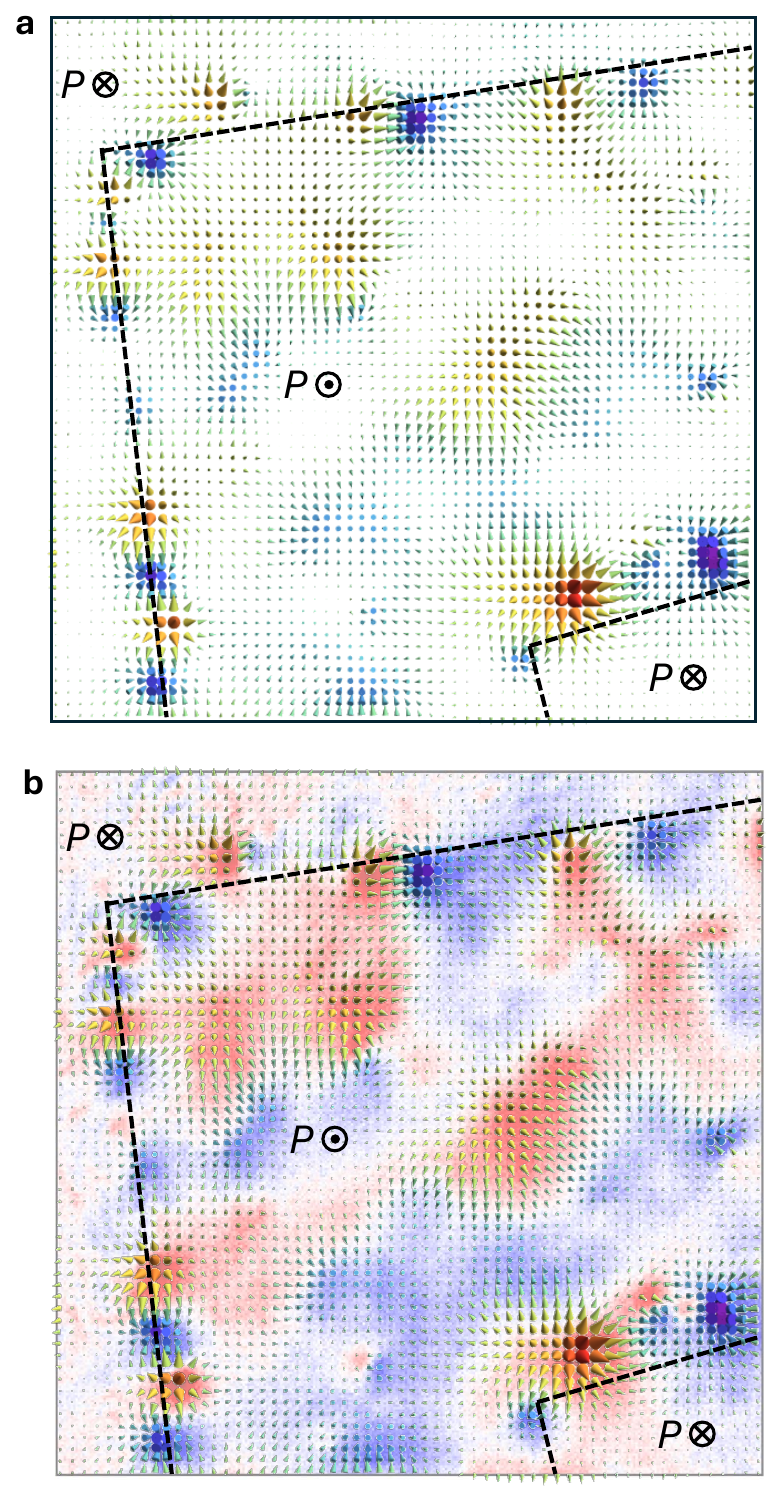}  \caption{(a) Un-normalized vector map of $B$ field in the upward poled region shown in the Fig. 4f of the main text. Panel (b) shows the same image overlaid on the full-$B$ image.}  \label{fig:s11}}
	\end{figure}

\clearpage

   	\begin{figure}[htp]
		\centering{
			{~}\hspace*{0cm}
			\includegraphics[scale=.58]{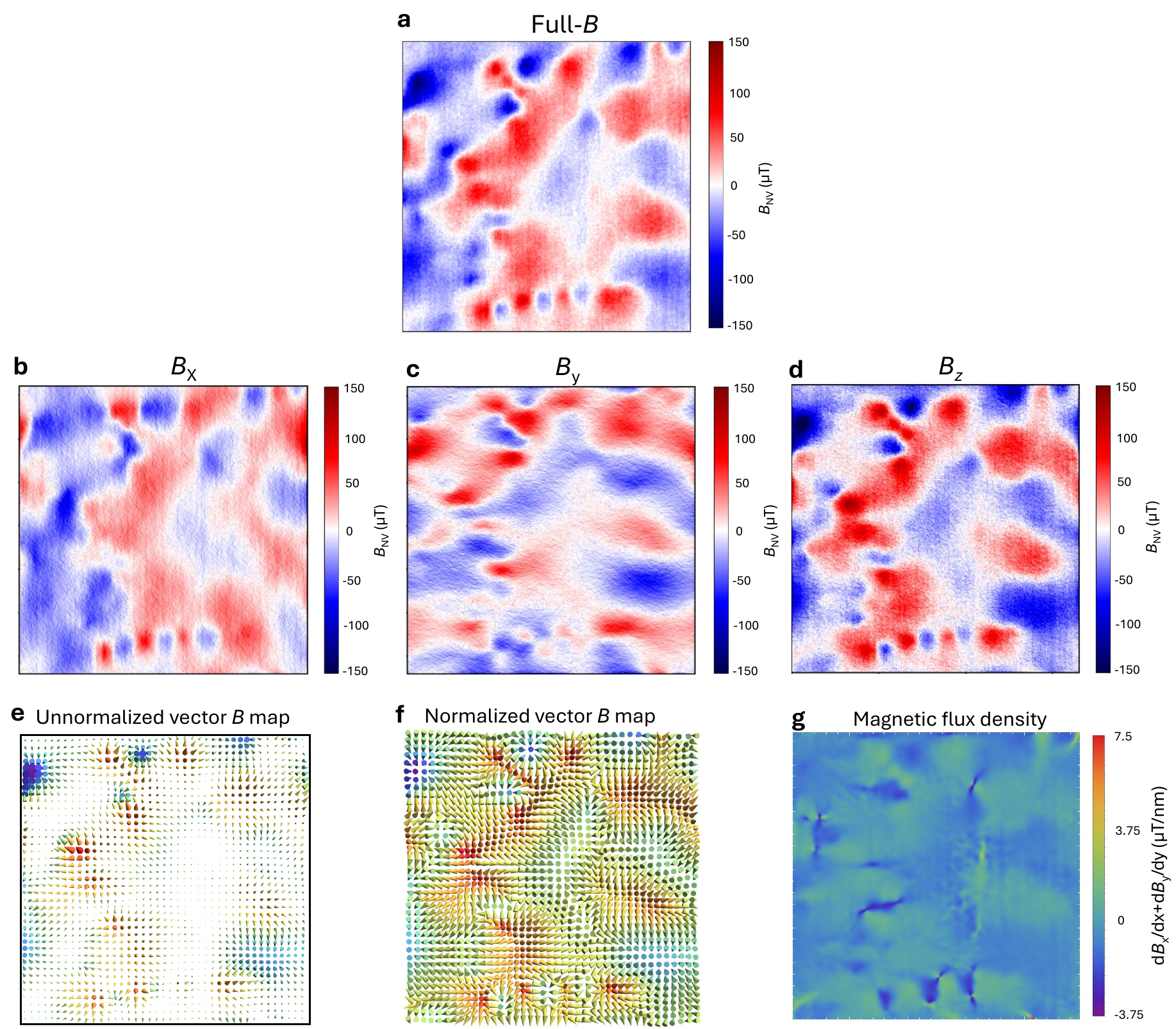}  \caption{ \textbf{Full-$B$ reconstruction in upward poled region on another sample:} For reproducibility, we have carried out Full-$B$ reconstruction in upward poled region on another sample. Panel (a) shows Full-$B$ image taken on another 1000 nm thick (111) oriented BiFeO$_3$ film. panels (b), (c) and (d) correspond to $x$, $y$ and $z$ component of stray magnetic field. (e) Vector $B$ map constructed by plotting the linear combination of $x$, $y$ and $z$ components at each pixel. To demonstrate the local orientation of $B$ field more clearly, panel (f) shows the same plot after normalizing each vector. (g) Magnetic flux density obtained from the plot shown in the panel (e).  \label{fig:s11}}}
	\end{figure}

\clearpage

    	\begin{figure}[htp]
		\centering{
			{~}\hspace*{-0.2cm}
			\includegraphics[scale=.72]{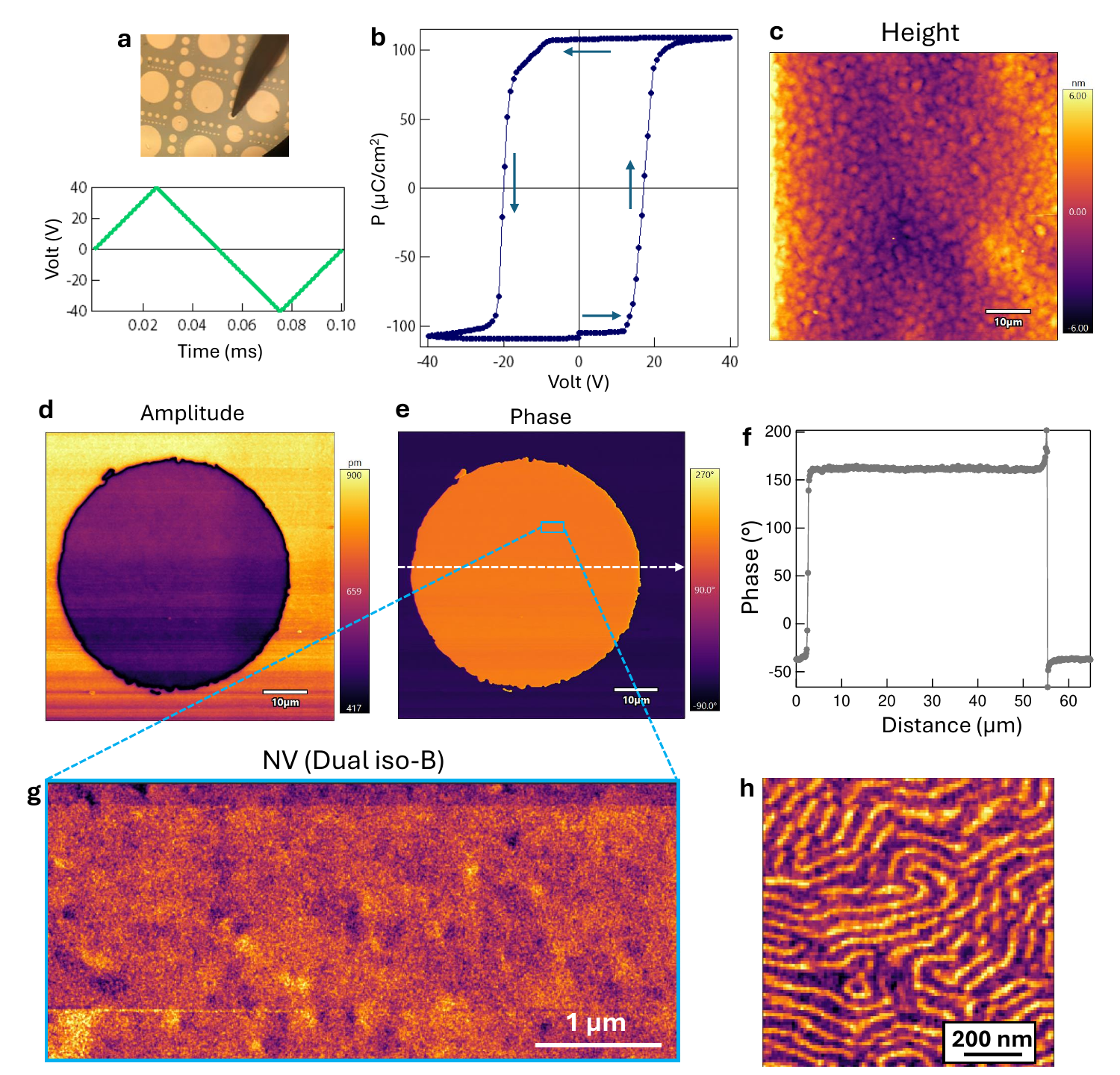}  \caption{ \textbf{NV image taken on region switched under top electrode:} (a)  (top panel) Optical image of the top electrode used to switch the {\bf{\textit{P}}} along  [111].  Black arrow like region is the electrical probe through which voltage was applied. For this experiment, gold electrode was used as it could be easily dissolved. This is required as the presence of electrode drastically reduces the NV signal from sample due to an increase in effective flying distance of NV$^-$ center. (bottom panel) Voltage ramping protocol utilized to switch the polarization.   Panel (b) shows the corresponding evolution of the polarization during this voltage ramping protocol. At the final stage, the remnant polarization is out of the plane i.e. along the [111] direction. After switching, gold electrodes were dissolved in potassium iodide solution (Methods).  Height profile of the surface after dissolving gold electrodes is shown in the panel (c). As evident pristine surface of BiFeO$_3$ is fully recovered thus allowing for NV measurements. Panels (d) and (e) show the PFM amplitude and phase respectively taken on one the regions switched electrode. As evident, region under electrode is completely switched and homogeneous over large scales. A phase change of 180 $^\circ$ warrants a complete switching of polarization from [$\bar{1}$$\bar{1}$$\bar{1}$] to [111]. Panel (g) shows the dual iso-$B$ image taken over 5 microns. As evident, the sample exhibits a homogeneous texture consistent with the results obtained on region switched with a conductive AFM tip. Please note that the outside regions still have glassy labyrinthine pattern (panel (h)) as discussed in the main text.} \label{fig:s12}}
	\end{figure}


\end{document}


\title{\Large \centering Supplementary Information\\ 
Morphogenesis of Spin Cycloids in a Non-collinear Antiferromagnet}

\vspace{1cm}
\makeatletter
\renewcommand{\maketitle}{\bgroup\setlength{\parindent}{10pt}
\begin{flushleft}
  \textbf{\@title}

  \@author
\end{flushleft}\egroup
}
\makeatother

\onecolumngrid

\vspace{20cm}
\setstretch{-1.0}
\date{}
\author{%
\textsf{\textbf{Shashank Kumar Ojha$^{1,*}$, Pratap Pal$^{2}$, Sergei Prokhorenko$^{3}$, Sajid Husain$^{4,*}$, Maya Ramesh$^{5}$, Peter Meisenheimer$^{6}$, Darrell G. Schlom$^{5,7,8}$, Paul Stevenson$^{9}$, Lucas Caretta$^{10}$ Yousra Nahas$^{3}$, Lane W. Martin$^{1,4,11,12}$, Laurent Bellaiche$^{3,13}$, Chang-Beom Eom$^{2}$, Ramamoorthy Ramesh$^{1,4,6,11,12,14,*}$\\}
$^{1}$Rice Advanced Materials Institute, Rice University, Houston, TX, 77005, USA\\
$^{2}$Department of Materials Science and Engineering, University of Wisconsin-Madison, Madison, WI 53706, USA\\
$^{3}$Smart Ferroic Materials Center, Physics Department and Institute for Nanoscience and Engineering, University of Arkansas, Fayetteville, Arkansas, USA\\
$^{4}$Materials Sciences Division, Lawrence Berkeley National Laboratory, Berkeley, CA, 94720, USA\\
$^{5}$Department of Materials Science and Engineering, Cornell University, Ithaca, NY, 14853, USA\\
$^{6}$Department of Materials Science and Engineering, University of California, Berkeley, CA, 94720, USA\\
$^{7}$Kavli Institute at Cornell for Nanoscale Science, Cornell University, Ithaca, NY, 14853, USA\\
$^{8}$Leibniz-nstitut fur Kristallzuchtung, Max-Born-Str. 2, 12489, Berlin, Germany\\
$^{9}$Department of Physics, Northeastern University, Boston, MA 02115, USA\\
$^{10}$School of Engineering, Brown University, Providence, RI, USA\\
$^{11}$Department of Materials Science and NanoEngineering, Rice University, Houston, TX, 77005, USA\\
$^{12}$Departments of Chemistry and Physics and Astronomy, Rice University, Houston, TX, 77005, USA\\
$^{13}$Department of Materials Science and Engineering, Tel Aviv University, Ramat Aviv, Tel Aviv 6997801, Israel\\
$^{14}$Department of Physics, University of California, Berkeley, CA, 94720, USA\\
{$^{*}$so37@rice.edu}\\
{$^{*}$rramesh@berkeley.edu}\\
{$^{*}$shusain@lbl.gov}\\
}}

\date{\today}

\maketitle

\setcounter{figure}{0}
\renewcommand{\figurename}{\textbf{Supplementary Figure}}
\newpage
\tableofcontents

\newpage

\setstretch{2}

\begin{flushleft}\section*{SUPPLEMENTARY FIGURE 1\\BiFeO$_3$ unit cell, structure of spin cycloid and spin density wave}
\end{flushleft}
	\begin{figure}[htp]
		\centering{
			{~}\hspace*{-0.2cm}
			\includegraphics[scale=.8]{Fig1.pdf}  \caption{ \textbf{BiFeO$_3$ unit cell, structure of spin cycloid and spin density wave (SDW)} (a) Two corner shared pseudocubic unit cells which constitute the rhombohedral lattice of BiFeO$_3$. Ferroelectric Polarization ({\bf{\textit{P}}}) is shown along the [111] direction. All crystallographic directions are in pseudocubic notations. In principle, {\bf{\textit{P}}} could point along any of the body diagonals.  $\bm{k}_1$, $\bm{k}_2$ and $\bm{k}_3$ denote the three equivalent directions for the propagation of antiferromagnetic spin cycloids in the bulk of BiFeO$_3$. For each $\bm{k}$, -$\bm{k}$ is the degenerate pair.  Orange arrows depict magnetic moment on Fe atoms.  Antiferrodistortive rotation of oxygen octahedra is shown with blue curved arrows. (b) Structure of spin cycloid along with spin density wave. Here the continuous cycloidal rotation of iron moments (orange arrows) in the {\bf{\textit{P}}}-{\bf{\textit{k}}} plane leads to the formation of spin cycloids.  } \label{fig:s1}}
	\end{figure}
 
\clearpage

\begin{flushleft}\section*{SUPPLEMENTARY FIGURE 2\\Crystal structure of (111) oriented BiFeO$_3$ FILM}
\end{flushleft}

	\begin{figure}[htp]
		\centering{
			{~}\hspace*{-0.2cm}
			\includegraphics[scale=.52]{Fig2.pdf}  \caption{ (a) X-ray diffraction pattern of a 1000 nm BiFeO$_3$/SrRuO$_3$/SrTiO$_3$ (111) heterostructure around (111) peak of STO. (b) Rocking curve of (111) BiFeO$_3$ peak. (c) Reciprocal space map taken around (111) peak. } \label{fig:s2}}
	\end{figure}

\clearpage

\begin{flushleft}\section*{SUPPLEMENTARY FIGURE 3\\Ferroelectric characterization of (111) BiFeO$_$ film}
\end{flushleft}

 	\begin{figure}[htp]
		\centering{
			{~}\hspace*{-0.2cm}
			\includegraphics[scale=.70]{Fig3.pdf}  \caption{ (a) Metal-ferroelectric-metal structure used for measuring polarization vs electric field hysteresis. Here SrRuO$_3$ layer serves as bottom electrode and platinum as top electrode. (b) Polarization vs electric field hysteresis loop measured on a 1000 nm BiFeO$_3$/SrRuO$_3$/SrTiO$_3$ (111) sample. Inset shows the voltage ramping protocol for measuring hysteresis. (c) Piezo force microscopy phases taken in vertical  channel. As evident, the variation in phase is very small over large scales demonstrating single ferroelectric domain. The as grown direction of the polarization is into the plane i.e. [$\bar{1}$$\bar{1}$$\bar{1}$]. This is also evident from the point-contact piezoelectric hysteresis (d) which exhibits a complete 180$^\circ$ phase reversal (from -90 $^\circ$ to +90$^\circ$) upon switching the electric field direction from [$\bar{1}$$\bar{1}$$\bar{1}$] to [111].} \label{fig:s3}}
	\end{figure}

\clearpage

\begin{flushleft}\section*{SUPPLEMENTARY NOTE 1\\ Calibrating the orientation of NV$^-$ center}
\end{flushleft}
   	\begin{figure}[htp]
		\centering{
			{~}\hspace*{-0.2cm}
			\includegraphics[scale=.50]{Fig8.pdf}  \caption{(a) Orientation of the NV$^-$ center in the laboratory frame of reference. (b) An optical image of a Ta(2 nm)/MgO/CoFeB(0.9 nm)/Ta(5 nm) sample with perpendicular magnetic anisotropy. Sky blue box shows the region where horizontal cuts of the magnetic profile across the edge of the sample was obtained. Full-$B$ scan of the same region is shown in the panel (c). (d) Line profile of the $B$ field (averaged over 50 linescans) across the edge. Solid green line shows the fitting of experimental data with a model described in the reference~\cite{PhysRevApplied.4.014003}. (e-g) Similar set of data obtained in another orientation (rotated 90 $^\circ$ with respect to the image in panel (b)).} \label{fig:s1}}
	\end{figure}
 
To make any quantitative estimations, it is essential to first calibrate the orientation of the NV$^-$  center relative to the laboratory reference frame. The NV$^-$  center in diamond aligns along the one of four equivalent $<111>$  crystallographic directions, which would theoretically correspond to a polar angle ($\theta$) of approximately 54.7 $^\circ$. To determine this, we followed the procedure outlined in previous studies~\cite{PhysRevApplied.4.014003,Meisenheimer2024}, where the magnetic field profile is recorded across the edge of a known ferromagnetic sample Ta(2 nm)/MgO/CoFeB(0.9 nm)/Ta(5 nm)  with perpendicular magnetic anisotropy. By measuring the $B$-field across edges of the sample in two different orientations of the sample and fitting the data to the analytical expressions provided in the reference~\cite{PhysRevApplied.4.014003}, we determined the $\theta$ to be 52 $^\circ$ and the $\phi$ to be 24 $^\circ$. These angles were then used to decompose the measured $B$-field, which is parallel to the NV$^-$ center, into its components along the x, y, and z axes in the laboratory frame, allowing us to construct vector maps of the magnetic field.

\clearpage

\begin{flushleft}\section*{SUPPLEMENTARY NOTE 2\\ Origin of stray $B$ field on (111) surface of BiFeO$_3$}
The full cycloidal magnetization profile of has three-dimensional texture; however, we can decompose this into two distinct contributions: the spiral-like texture of the spins (which likes in the plane defined by the polarization $\vec{P}$ and the cycloid propagation direction $\vec{k}$), and the spin density wave formed by the canted moments of these spins, which likes in the plane perpendicular to $\vec{P}$. For clarity, we will refer to the former as the ``spiral'' contribution to avoid confusion with the net cycloidal texture.

In (001) BiFeO$_3$, the spiral contribution is near-zero because adjacent lattice sites at the surface layer (and each layer below) have oppositely-directed spins, resulting in almost no net surface magnetization. However, the spin density wave is not cancelled and yields a net stray field.

In (111) BiFeO$_3$, this is not the case. If we define a coordinate system where $z$ is parallel to $\vec{P}$ and $x$ is parallel to $\vec{k}$, we can write the following for the SDW:
\begin{align}
    \vec{M}_{SDW} & = A \cos(\vec{k} \cdot \vec{r})(\vec{e}_y)  \\
    & = A \cos(|k|x)(\vec{e}_y)
\end{align}

In reciprocal space, this yields:
\begin{align}
    \tilde{M_x}(q) & = 0 \\
    \tilde{M_y}(q) & = A \delta(q_y) \bigg(\delta(q_x-k)  + \delta(q_x + k)  \bigg) \\ 
    \tilde{M_z}(q) & = 0 
\end{align}

Combined with the Fourier-domain approach to calculating the stray field:
\begin{gather}
    \begin{pmatrix} \tilde{B}_x \\ \tilde{B}_y  \\ \tilde{B}_z \end{pmatrix} = -\frac{\mu_0}{2} e^{qz}
    \begin{pmatrix}
        q_x^2/q & q_x q_y/q & i q_x \\ q_x q_y/q & q_y^2/q & i q_y \\ i q_x & i q_y & -q
    \end{pmatrix}
    \begin{pmatrix}
        \tilde{M}_x \\ \tilde{M}_y  \\ \tilde{M}_z
    \end{pmatrix}
\end{gather}
this yields $\vec{B}=0$ for the SDW component.

We can calculate the stray field generated by the spiral ordering from a single monolayer of spins using:
\begin{equation}
    \vec{M}{spiral} = m_{Fe}
    \begin{pmatrix}
        \cos(kx) \\  0 \\ \sin(kx)
    \end{pmatrix}
\end{equation}
where we find that now the generated stray field is \textit{not} zero. 

However, we must address the question of cancellation of these magnetic moments from the antiferromagnetic ordering. If we define the stray field generated by a monolayer of our [111] orientation as $\vec{B}$, then the layer below this contributes a stray field of $\exp(-|k|a)\vec{B}$, where $|k|$ is the cycloid wavenumber and $a$ is the layer spacing. This allows us to write the net observed field as:
\begin{align}
    \vec{B}_{net} &  = \vec{B} \sum_{n=0}^{\infty} (-1)^{n}\exp(-nka) \\
    & = \vec{B}\frac{1}{1 + \exp(-ka)}
\end{align}
which for $k=\frac{2\pi}{65\mathrm{nm}}$ and $a=\sqrt{3}\times 0.395\mathrm{nm}$ gives $\vec{B}_{net} \approx 0.5 \vec{B}$.

Thus, while the NV signal observed in (111) and (001) BiFeO$_3$ appears similar, the origin of the detected stray field is distinct in each case, potentially allowing different terms in the BiFeO$_3$ Hamiltonian to be probed using different substrate orientations.
\end{flushleft}

\bibliographystyle{naturemag}
	\bibliography{111_turing_BFO.bib}